\documentclass[prl,aps,twocolumn,superscriptaddress,preprintnumbers,nopacs,floatfix,amsmath,amssymb]{revtex4}

\usepackage{graphicx}% Include figure files
\usepackage{dcolumn}% Align table columns on decimal point
\usepackage{bm}% bold math

\usepackage{amsmath}
\usepackage{graphicx}
\usepackage{amssymb}
\usepackage{mathrsfs}
\usepackage{bbm}
\usepackage{epsfig}
\newcommand{\be}{\begin{eqnarray}}
\newcommand{\ee}{\end{eqnarray}}

\begin{document}

%\pagenumbering{empty}
%\begin{titlepage}
%
\title{Virtual reality based  approach to 
protein \\ heavy-atom structure reconstruction} 

%\vskip 5.0cm

\author{Xubiao Peng}
\email{xubiaopeng@gmail.com}
\affiliation{Department of Physics and Astronomy, Uppsala University,
P.O. Box 803, S-75108, Uppsala, Sweden}

\author{Alireza Chenani}
\email{achenani@gmail.com}
\affiliation{Department of Physics and Astronomy, Uppsala University,
P.O. Box 803, S-75108, Uppsala, Sweden}

\author{Shuangwei Hu}
\email{hushuangwei@gmail.com}
\affiliation{Department of Physics and Astronomy, Uppsala University,
P.O. Box 803, S-75108, Uppsala, Sweden}

\author{Yifan Zhou}  
\email{ahan.zhou@gmail.com}
\affiliation{Department of Biomedicine 
Faculty of Medicine and Dentistry, UIB
Jonas Lies Vei 91, NO-5009 Bergen, Norway}

\author{Antti J. Niemi}
\email{Antti.Niemi@physics.uu.se}
\affiliation{ Laboratoire de Mathematiques et Physique Theorique
CNRS UMR 6083, F\'ed\'eration Denis Poisson, Universit\'e de Tours,
Parc de Grandmont, F37200, Tours, France}
\affiliation{Department of Physics and Astronomy, Uppsala University,
P.O. Box 803, S-75108, Uppsala, Sweden}

\begin{abstract}

{\bf Background}
A commonly recurring problem in structural protein studies, is  the determination of  all heavy atom 
positions from the knowledge of
the  central $\alpha$-carbon coordinates.  

\vskip 0.2cm
{\bf Results }
 We employ advances in virtual reality
to address the problem.  
The outcome is a  3D visualisation based technique 
where all the heavy backbone and side chain atoms are treated on equal footing,  in 
terms of the C$_\alpha$ coordinates. Each heavy atom can be visualised on
the surfaces of the different two-spheres, that are centered at the other heavy backbone and side chain atoms. 
In particular, the rotamers are 
visible as clusters which display strong dependence  
on the underlying backbone secondary structure. 
 
 \vskip 0.2cm
 {\bf Conclusions} 
Our method  easily detects those atoms in a crystallographic protein structure which have been been likely misplaced. 
Our approach forms a basis for the development of a new generation, visualisation based 
side chain construction, validation and refinement tools.  The heavy atom positions are identified 
in a manner which accounts for the secondary structure environment,  leading to improved accuracy over existing methods. 

\vskip 1.0cm

\end{abstract}

%\date{\today}

\maketitle

Protein structure validation methods like MolProbity \cite{Chen-2010} and Procheck  
\cite{Laskowski-1993} help crystallographers to find and fix potential problems that are
incurred during fitting and refinement. These methods are commonly based 
on  {\it a priori} chemical knowledge and utilize various
well tested and broadly accepted stereochemical paradigms.
Likewise,  template based structure prediction and analysis
packages  \cite{Qu-2009} and  molecular dynamics 
force fields  \cite{Freddolino-2010} are customarily built on such paradigms.
Among these, the
Ramachandran map \cite{Ramachandran-1963}, \cite{Carugo-2013} has a central r\^ole.
It is widely deployed both to various analyzes of  the protein structures, and 
as a tool in protein visualization.  
The Ramachandran map describes the statistical distribution of the two dihedral 
angles $\phi$ and $\psi$ that are adjacent to the C$_\alpha$ carbons
along the protein backbone. A comparison between 
the observed values of the individual dihedrals in a given protein with the
statistical distribution of the Ramachandran map is an appraised
method to validate the backbone geometry. 

In  the case of side chain atoms, visual analysis methods alike
the Ramachandran map have been introduced. For example, 
the Janin map \cite{Janin-1978} can be used to compare observed side chain
dihedrals such as $\mathcal X_1$ and $\mathcal X_2$ in a given protein,
against their statistical distribution, in a manner which is analogous to the 
Ramachandran map. 
Crystallographic refinement and validation programs like 
Phenix \cite{Adams-2010}, Refmac \cite{Murshudov-1997} and others,
often utilize the statistical data obtained 
from the Engh and Huber library \cite{Engh-1991}, \cite{Engh-2001}. This library is built using small molecular 
structures that have been determined with a very high resolution. 
At the level of entire proteins, side chain restraints are commonly derived
from  analysis of high resolution crystallographic structures \cite{Ponder-1987},
\cite{Dunbrack-2002} in Protein Data Bank (PDB) \cite{Berman-2000}. 
A backbone independent rotamer library \cite{Lovell-2000}
makes no reference to backbone conformation. But the possibility that
the side-chain rotamer population depends on  the local protein backbone 
conformation, was  considered already by Chandrasekaran and 
Ramachandran \cite{Chandrasekaran-1970}. Subsequently both secondary 
structure dependent \cite{Schrauber-1993}, see also \cite{Janin-1978} and \cite{Lovell-2000}, and 
backbone dependent rotamer libraries  \cite{Dunbrack-1993}, \cite{Shapovalov-2011}
have been developed. The information content in the secondary structure 
dependent libraries and the backbone independent libraries essentially 
coincide \cite{Dunbrack-2002}. Both kind of libraries are used extensively during crystallographic
protein structure model building  and refinement. But for the prediction 
of side-chain conformations  for example 
in the case of  homology modeling and protein design, there can be an 
advantage to use the more revealing 
backbone dependent rotamer libraries.

In x-ray crystallographical protein structure experiments, the skeletonization of the electron 
density map is a common technique to interpret the data and to
build the initial model  \cite{Jones-1991}. 
The C$_\alpha$ atoms are located at the branch points between the 
backbone and the side chain, and as such 
they are subject  to relatively stringent stereochemical constraints; this is the reason why the model building often 
starts with the initial identification of the skeletal C$_\alpha$ trace.
The central r\^ole of the C$_\alpha$ atoms is widely
exploited in structural classification schemes such as  CATH  \cite{Sillitoe-2013} and SCOP \cite{Murzin-1995}, 
in various threading \cite{Roy-2010} and homology \cite{Schwede-2003} modeling techniques \cite{Zhang-2009},  in {\it de novo}  approaches \cite{Dill-2007},
and in the development of coarse grained energy functions for folding prediction \cite{Scheraga-2007}.
As a consequence the so-called C$_\alpha$-trace problem
has become the subject of extensive investigations  \cite{Holm-1991,DePristo-2003,Lovell-2003,Rotkiewicz-2008,Li-2009}. 
The resolution of the problem would consist of 
an accurate main chain and/or all-atom model of the folded protein from 
the knowledge of the positions of the central C$_\alpha$ atoms only.
Both knowledge-based approaches such and  MAXSPROUT \cite{Holm-1991}  and {\it de novo} methods
including PULCHRA \cite{Rotkiewicz-2008} and  REMO \cite{Li-2009} have been developed, to try and
resolve the C$_\alpha$ trace problem.  
In the case of the backbone atoms,
the geometric algorithm introduced  by  Purisima and Scheraga \cite{Purisima-1984}, or 
some variant thereof, is commonly utilized in these approaches. For the side chain atoms, 
most approaches to the C$_\alpha$ trace problem rely either on a statistical
or on a conformer rotamer library in combination with steric constraints,  complemented by an analysis
which is based on diverse scoring functions. For the final fine-tuning of the model, 
all-atom molecular dynamics simulations can also be utilized.

In the present article we introduce and develop 
new generation visualization techniques that we hope will become a beneficial
component  in protein structure analysis, refinement and validation. In line with the concept of the
C$_\alpha$ trace problem we deploy only a geometry that is determined solely in terms of the
C$_\alpha$ coordinates. The output we aim at, is a  3D "what-you-see-is-what-you-have" type 
visual map of the statistically preferred all-atom model, calculable in terms of the C$_\alpha$ coordinates. 
As such, our approach should have value
for example  during 
the construction and validation of the initial backbone and
all-atom models of a crystallographic protein structure.
%
%method, to statistically resolve the all-atom structure  of a protein
%from the knowledge of the C$_\alpha$ coordinates in crystallographic x-ray structures. 

Our approach is based on developments in three dimensional visualization and virtual reality,
that have taken place mainly after the Ramachandran map was introduced. In lieu of the backbone
dihedral angles that appear as coordinates in the Ramachandran map and
correspond to a toroidal topology, we employ 
the geometry of virtual two-spheres that surround each heavy atom. 
We visually describe all the higher level heavy backbone and side chain atoms on the surface of the sphere,
level-by-level along the backbone and side chains, 
exactly in the manner how they are seen by an imaginary, geometrically determined and C$_\alpha$ based
miniature observer who roller-coasts along the backbone and climbs up the side chains, while
proceeding from one C$_\alpha$ atom to the next.  At the location of each C$_\alpha$ our virtual 
observer orients herself
consistently according to the purely geometrically determined C$_\alpha$ based
discrete Frenet frames \cite{Hu-2011,Lundgren-2012a}.  
Thus the visualization depends only on the C$_\alpha$ coordinates, there is no reference to the 
other atoms in the initialization of  the construction. The other atoms - including subsequent C$_\alpha$ atoms
along the backbone chain - are all mapped 
on the surface of a sphere that surrounds the observer, as if these 
atoms were stars in the sky. 

At each C$_\alpha$ atom,
the construction proceeds along the ensuing side chain, until the position of
all heavy atoms have been determined.
As such our maps provide  a purely geometric and equitable, direct visual information on the statistically 
expected all-atom structure in a given protein.  

The method we describe in this article, can form a basis for the future development of a 
novel approach to the C$_\alpha$ trace problem. Unlike the existing approaches such as 
MAXSPROUT \cite{Holm-1991},  PULCHRA \cite{Rotkiewicz-2008} and  
REMO \cite{Li-2009} the method we envision accounts for the secondary structure dependence in
the heavy atom positions, which we here reveal. A secondary-structure dependent method to resolve
the C$_\alpha$ trace problem should lead to
an improved accuracy in the heavy atom positions, in terms of the C$_\alpha$
coordinates. The present article  is a proof-of-concept.

\section{Method and Results}

\subsection{C$_\alpha$ based Frenet frames}

\begin{figure}[h]
        \centering
                \includegraphics[width=0.45\textwidth]{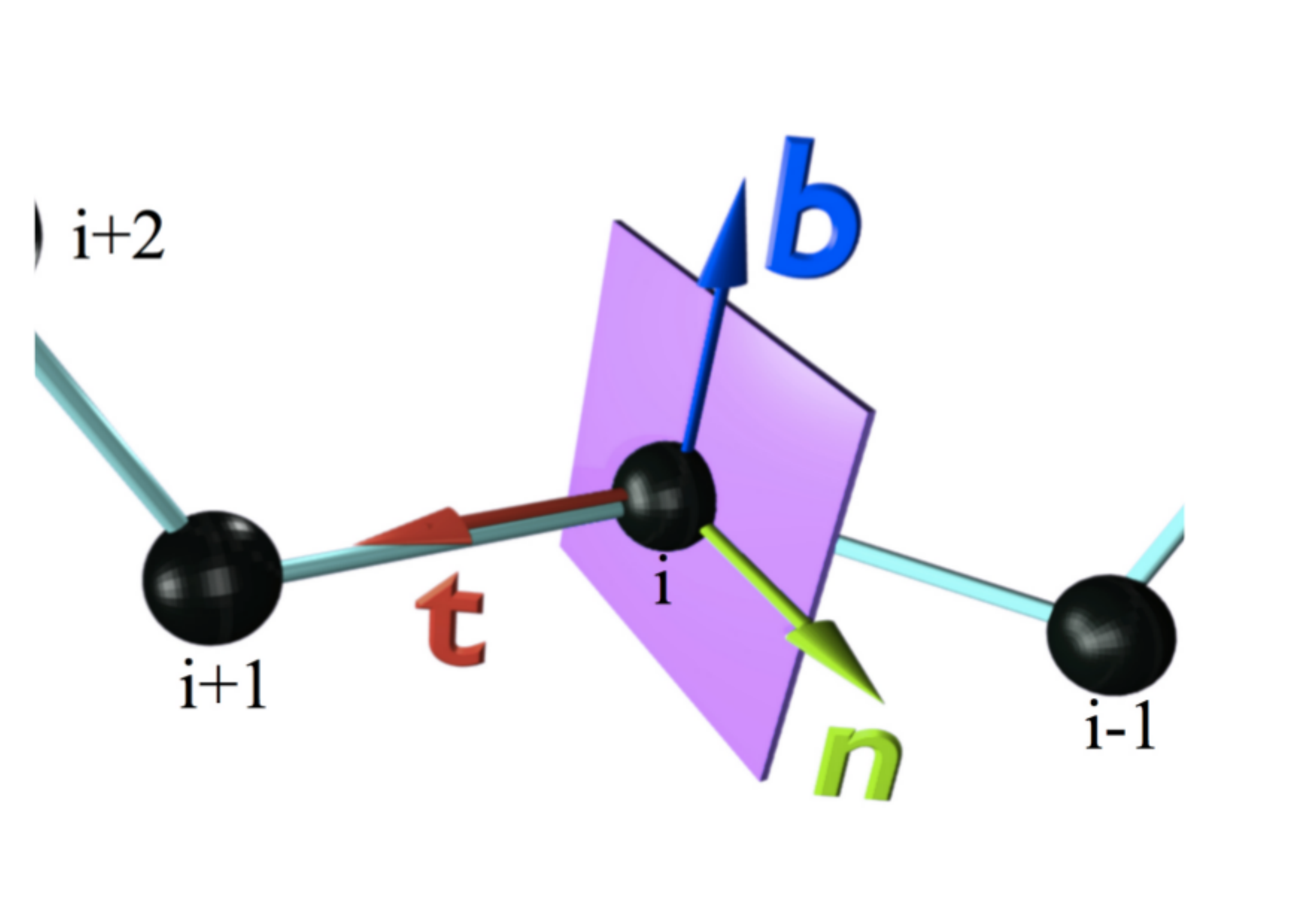}
        \caption{{ 
     (Color online) Discrete Frenet frame vectors (\ref{t}), (\ref{b}) and (\ref{n}).
       }}
       \label{fig1}
\end{figure}

\begin{figure}[h]
        \centering
                \includegraphics[width=0.45\textwidth]{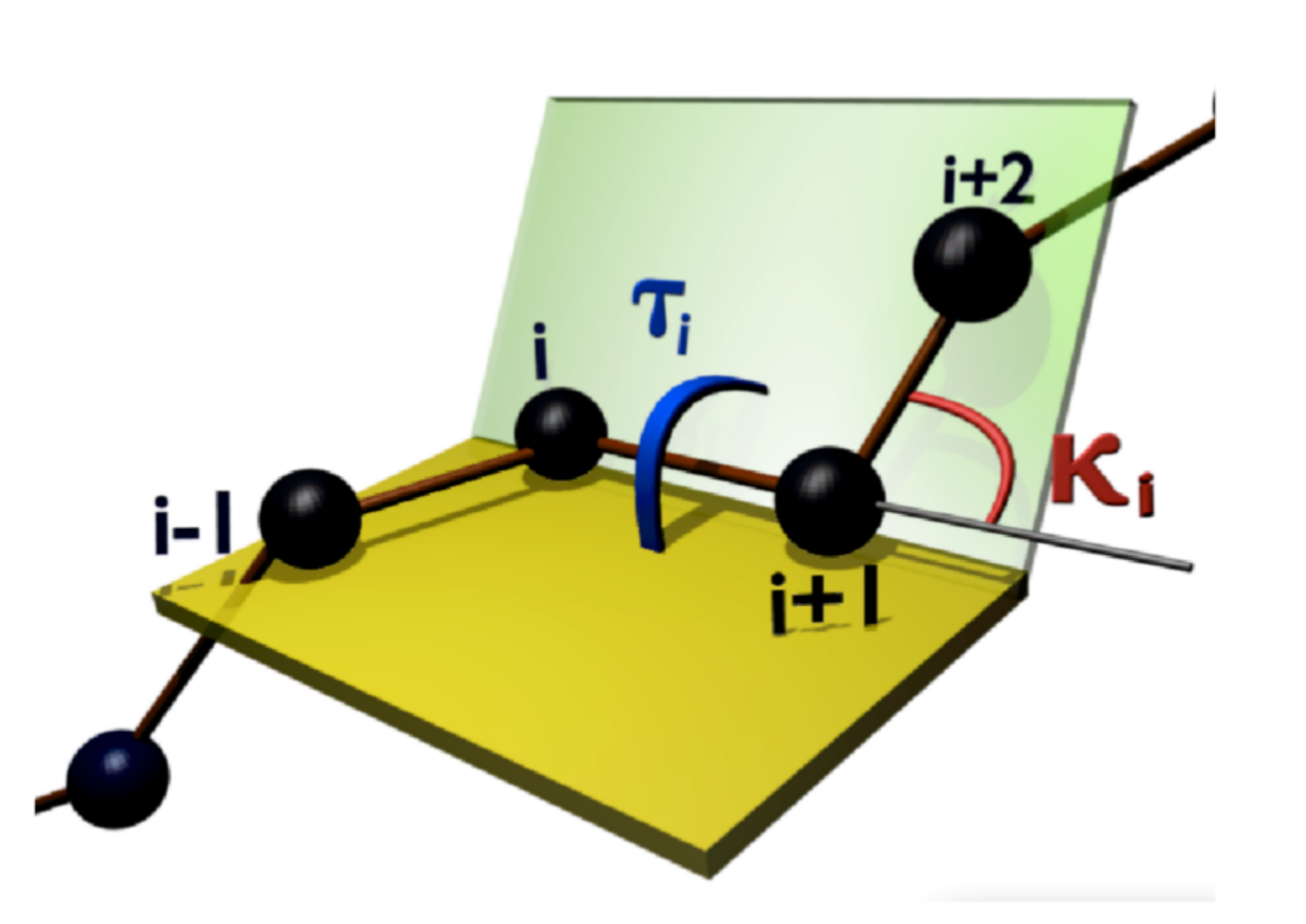}
        \caption{{ 
     (Color online) Geometry of bond ($\kappa_i$)  and torsion  ($\tau_i$)  angles (\ref{kappa}) and (\ref{tau}).
       }}
       \label{fig2}
\end{figure}

\begin{figure}[h]
        \centering
                \includegraphics[width=0.45\textwidth]{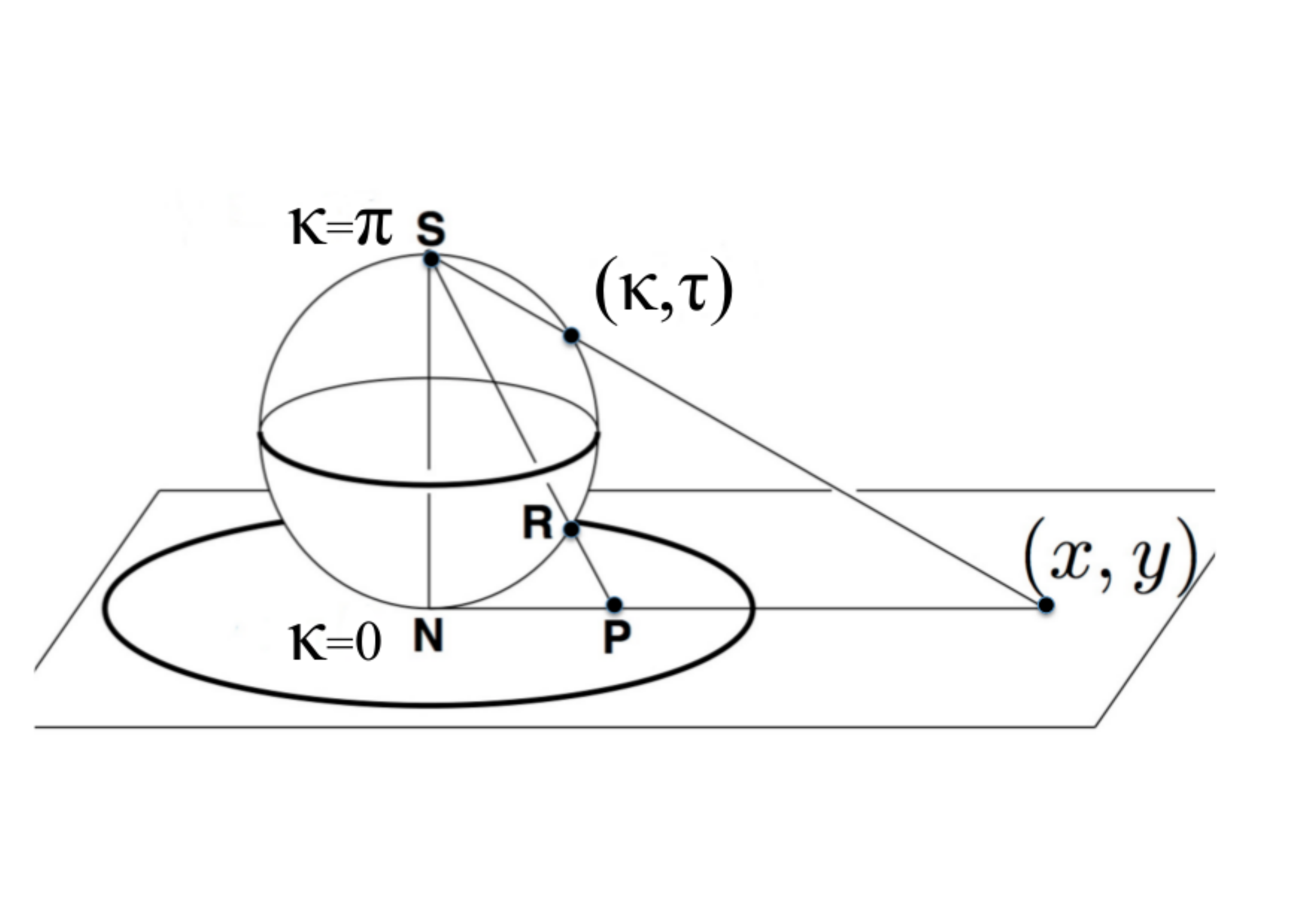}
        \caption{{ 
     (Color online) Geometry of bond ($\kappa_i$)  and torsion  ($\tau_i$)  angles (\ref{kappa}) and (\ref{tau}).
       }}
       \label{fig3}
\end{figure}

Let $\mathbf r_i$ ($i=1,...,N$) be the coordinates  of the  C$_\alpha$ atoms.
The counting starts from the N terminus.  At each $\mathbf r_i$ we introduce the orthonormal,
right-handed, discrete Frenet frame ($\mathbf t_i, \mathbf n_i , \mathbf b_i$) \cite{Hu-2011}. 
As shown in figure \ref{fig1} 
%\marginpar{Fig. \ref{fig1}}
%%%%%%%%%%%%%%%%%%%%%%%%%%%%%%%%%%%%%%%%%%%%%
%
%
%
%
%
%figure -1
%
%
%
%
%%%%%%%%%%%%%%%%%%%%%%%%%%%%%%%%%%%%%%%%%%%%%
%
%
the tangent vector $\mathbf t$
points from the center of the $i^{th}$ central carbon towards the center of the $(i+1)^{st}$ central carbon,
\begin{equation}
\mathbf t_i \ = \ \frac{\mathbf r_{i+1} - \mathbf r_i}{|\mathbf r_{i+1} - \mathbf r_i|} 
\label{t}
\end{equation}
The binormal vector is
\begin{equation}
\hskip -0.45cm \mathbf b_i \ = \ \frac{ \mathbf t_{i-1} \times \mathbf t_i }{ |  \mathbf t_{i-1} \times \mathbf t_i |}
\label{b}
\end{equation}
The normal vector is
\begin{equation}
\hskip -0.9cm \mathbf n_i \ = \ \mathbf b_i \times \mathbf t_i
\label{n}
\end{equation}
%
%
%
%
%       %%%%%%%%%%%%%%%%%%%%%%%%%%%%%%
%
%
%
%
We also introduce the virtual C$_\alpha$ backbone bond ($\kappa$) and torsion ($\tau$) 
angles, shown in figure \ref{fig2}.
%\marginpar{Fig. \ref{fig2}}
%%%%%%%%%%%%%%%%%%%%%%%%%%%%%%%%%%%%%%%%%%%%%
%
%
%
%
%
%figure -2
%
%
%
%
%%%%%%%%%%%%%%%%%%%%%%%%%%%%%%%%%%%%%%%%%%%%%
%
%
These angles are computed as follows,
\begin{equation}
\cos\kappa_{i+1} \ = \ \mathbf t_{i+1} \cdot \mathbf t_i
\label{kappa}
\end{equation}
\begin{equation}
\cos\tau_{i+1} \ = \ \mathbf b_{i+1} \cdot \mathbf b_i
\label{tau}
\end{equation}
We identify the bond angle $\kappa \in [0,\pi]$ with the latitude angle  of a two-sphere
which is centered at the C$_\alpha$ carbon. We orient the sphere so that  the 
north-pole where $\kappa=0$ is in the direction of $\mathbf t$.  
The torsion angle $\tau\in [-\pi,\pi)$ is the longitudinal angle. It is defined so that $\tau = 0$ 
on the great circle that passes both through the north pole and through the tip of the normal vector 
$\mathbf n$. The longitude angle increases towards the counterclockwise direction 
around the  vector $\mathbf t$.
Additional visual gain can be obtained, by stereographic projection 
of the sphere onto the plane.  The standard
stereographic projection  from the south-pole of the sphere to the plane with coordinates ($x,y$) 
is given by
\begin{equation}
x+iy \ \equiv \ \sqrt{ x^2 + y^2} \, e^{i\tau} \ = \ \tan\left( \kappa/2 \right) \, e^{i\tau}
\label{stereo}
\end{equation}
This maps the north-pole where $\kappa=0$ to the origin ($x,y$)$=$($0,0$). The south-pole where 
$\kappa=\pi$ is sent to infinity; see figure \ref{fig3}
%\marginpar{Fig. \ref{fig3}}. 
%%%%%%%%%%%%%%%%%%%%%%%%%%%%%%%%%%%%%%%%%%%%%
%
%
%
%
%
%figure -3
%
%
%
%
%%%%%%%%%%%%%%%%%%%%%%%%%%%%%%%%%%%%%%%%%%%%%
%
%
The visual effects  
can be further enhanced by  sending
\begin{equation}
\kappa \ \to \ f(\kappa) 
\label{ste2}
\end{equation}
where $f(\kappa)$ is a properly chosen function of the latitude angle $\kappa$. Various different choices of 
$f(\kappa)$  will be considered in the sequel.

\subsection{The C$_\alpha$ map}

We first describe, how to visually characterize the C$_\alpha$ trace in terms of the C$_\alpha$ based
Frenet frames (\ref{t})-(\ref{n}). We introduce the concept of a virtual 
miniature observer who roller-coasts the backbone by moving  between the C$_\alpha$ atoms.
At the location of each C$_\alpha$ the observer has an orientation that is determined
by the Frenet frames (\ref{t})-(\ref{n}). The base of the $i^{th}$ tangent vector {$\mathbf t_i$ is  
at the position $\mathbf r_{i}$.  The tip of $\mathbf t_i$  
is a point on the surface  of the sphere ($\kappa,\tau$) that surrounds the  observer; it points 
towards the north-pole. The vectors $\mathbf n_i$ and $\mathbf b_i$ determine the orientation of the sphere,
these vectors define 
a frame on the normal plane to the backbone trajectory, as shown in figure \ref{fig1}. 
The observer uses the sphere to  construct a map of the  various atoms in the protein chain. 
She identifies them  as points on the surface of the two-sphere that surrounds her, 
as if the atoms were stars in the sky.

The observer constructs  the C$_\alpha$ backbone map as follows \cite{Lundgren-2012a}.
She first translates the center of the  sphere from the location of the $i^{th}$ C$_\alpha$,  
all the way to the location of the $(i+1)^{th}$ C$_\alpha$,  without introducing any rotation 
of the sphere, with respect to the $i^{th}$ Frenet frames.  She then 
identifies the direction of $\mathbf t_{i+1}$, {\it i.e.} the 
direction towards the site $\mathbf r_{i+2}$ to which she proceeds from the next C$_\alpha$ carbon,
as a point on the surface of the sphere. This determines  the corresponding coordinates ($\kappa_i, \tau_i$).  
After this, she re-defines her orientation to match the Frenet framing at the $(i+1)^{th}$ central carbon, and proceeds
in the same manner.
The ensuing map, over the entire backbone, gives an instruction to the observer at each
point $\mathbf r_i$,  how to turn at site $\mathbf r_{i+1}$, to 
reach  the  $(i+2)^{th}$ C$_\alpha$ carbon at the point $\mathbf r_{i+2}$.   

In figure \ref{fig4} (top)
%\marginpar{Fig. \ref{fig4}}
%%%%%%%%%%%%%%%%%%%%%%%%%%%%%%%%%%%%%%%%%%%%%
%
%
%
%
%
%figure -4
\begin{figure}[h]
        \centering
                \includegraphics[width=0.45\textwidth]{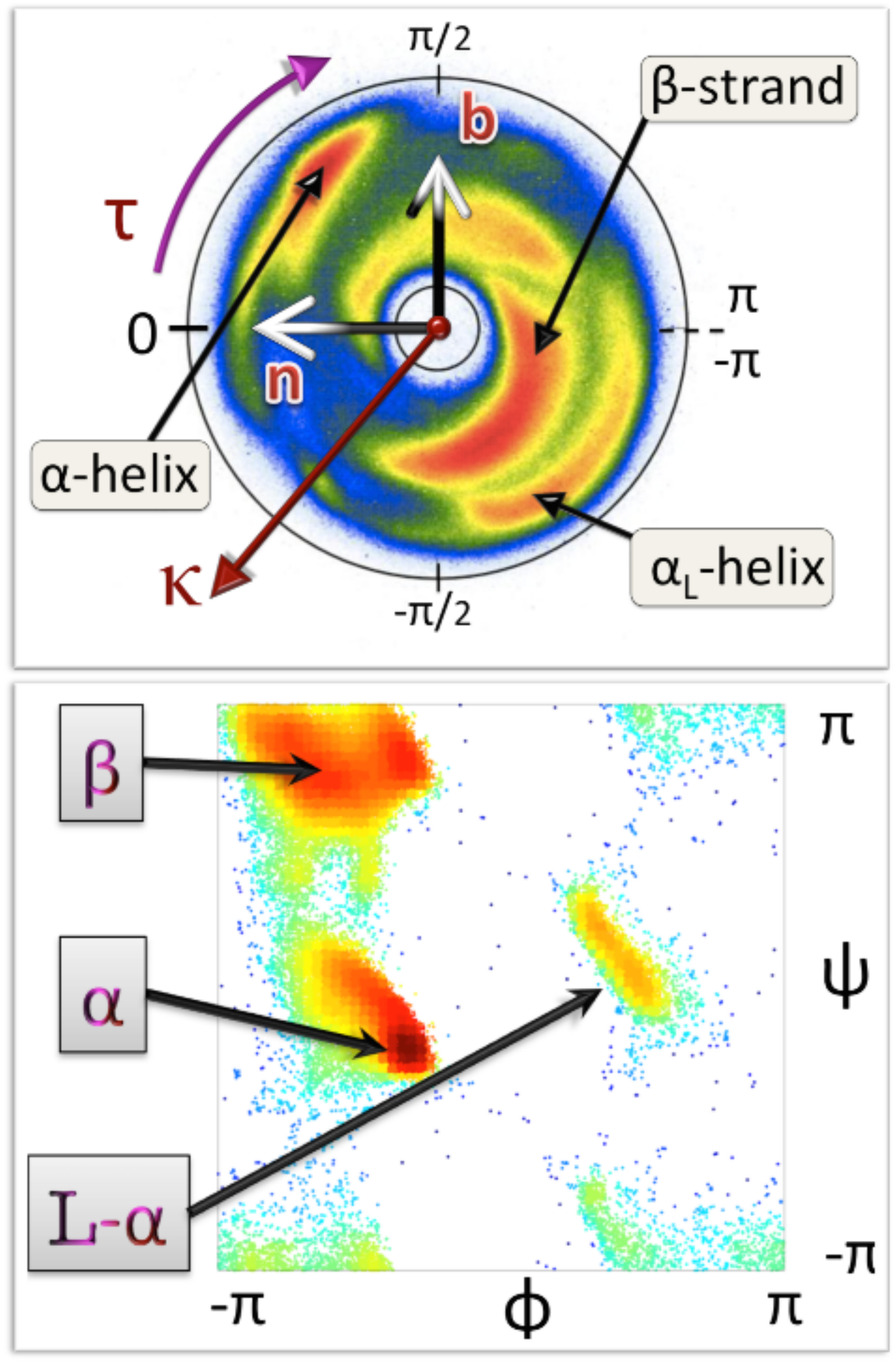}
        \caption{{ 
     (Color online) Top: The stereographically projected Frenet frame map of backbone C$_\alpha$ atoms, with major 
     secondary
structures identified. Also shown is the directions of the  Frenet frame normal vector $\mathbf n$; the  vector $\mathbf t$
corresponds to the red circle at the center, and it points away from the viewer. The map is constructed using all
PDB structures that have been measured with better than 2.0 \AA~ resolution. Bottom: Standard Ramachandran map,
constructed using our 1.0 \AA~ resolution PDB subset. Major secondary structures have been identified. 
       }}
       \label{fig4}
\end{figure}
%
%
%
%
%%%%%%%%%%%%%%%%%%%%%%%%%%%%%%%%%%%%%%%%%%%%%
%
%
we show the  C$_\alpha$ Frenet frame backbone map. It describes the statistical distribution that we obtain
when we plot all PDB structures which have been 
measured with better than 2.0 \AA~
resolution, and using the stereographic projection (\ref{stereo}); for statistical clarity we prefer to use 
here a more extended subset of PDB, than our canonical 1.0 \AA~subset, which we shall use in the remainder of the 
present article. Here the difference is minor.

For our observer, who always fixes her gaze position towards the north-pole 
of the surrounding two-sphere at each C$_\alpha$  {\it i.e.}  towards the red dot at the center of the annulus,  
the color intensity in this map 
reveals the probability of the direction at position $\mathbf r_i$, where 
the observer will turns at the next C$_\alpha$ carbon, when she moves 
from $\mathbf r_{i+1}$ to $\mathbf r_{i+2}$. In this way, the map is in a direct visual correspondence with
the way how the Frenet frame observer perceives the backbone geometry.  
We note that the probability distribution concentrates within an 
annulus, roughly between the latitude angle values $\kappa \sim1$ and $\kappa \sim 3/2$.
The exterior of the annulus is a sterically excluded region while  the entire interior is in principle sterically 
allowed but  not occupied in the case of folded proteins.
In the figure we identify four major secondary structure regions, according to the PDB classification.
These are $\alpha$-helices, $\beta$-strands, left-handed $\alpha$-helices
and loops. In this article we will use this rudimentary level PDB classification thorough. 

We note that the visualization in  
figure \ref{fig4} (top) resembles the Newman projection of stereochemistry:
The vector $\mathbf t_{i} $ which is denoted
by the red dot at the center of the figure, points along the backbone
from the promixal C$_\alpha$
at  $\mathbf r_{i}$ towards the
distal C$_\alpha$ at $\mathbf r_{i+1}$. This 
convention will be used thorough the present article.

When we surround C$_\alpha$  with an imaginary two-sphere, with C$_\alpha$ at the origin,
we may choose the radius of the sphere to coincide with the (average) 
covalent bond length value \cite{Lundgren-2012a} which is 3.8 \AA~in the case of C$_\alpha$ atoms, excluding the
{\it cis-}proline . Since the variations in the covalent bond lengths are in general
minor, in this article we do not account for deviations
in covalent bond lengths from their ideal values.

For comparison, we also show in figure \ref{fig4} (bottom)
the standard Ramachandran map. The sterically allowed and excluded regions are now intertwined,
while the allowed regions are more localized than in figure \ref{fig4} (top).
We point out that the map in figure \ref{fig4} (top) provides 
non-local information on the backbone geometry, it extends over several peptide units, and tells the 
miniature observer where the backbone turns at the next C$_\alpha$.  
As such it goes beyond the regime of the Ramachandran map, which  is 
localized to a single C$_\alpha$ carbon and does not provide direct
information how the backbone proceeds:  The
two Ramachandran angles $\phi$ and $\psi$ are dihedrals for a given C$_\alpha$,
around the N-C$_\alpha$ and C$_\alpha$-C covalent bonds. These angles to not 
furnish information
about neighboring peptide groups. 
%We also  remind  that the topology of the
%Ramachandran map is different, instead of a sphere it describes a torus. 

\subsection{Backbone heavy atoms}

Consider our imaginary miniature observer, 
located at the position of a C$_\alpha$ atom and oriented according to the discrete
Frenet frames. 
She observes and records 
the backbone heavy atoms N, C and the side-chain C$_\beta$
that are covalently bonded to a given C$_\alpha$, and the O in the peptide plane that precedes
C$_\alpha$. 
In figures \ref{fig5} a)-d)
%\marginpar{Fig. \ref{fig5}}
%%%%%%%%%%%%%%%%%%%%%%%%%%%%%%%%%%%%%%%%%%%%%
%
%
%
%
%
%figure -5
\begin{figure}[h]
        \centering
                \includegraphics[width=0.45\textwidth]{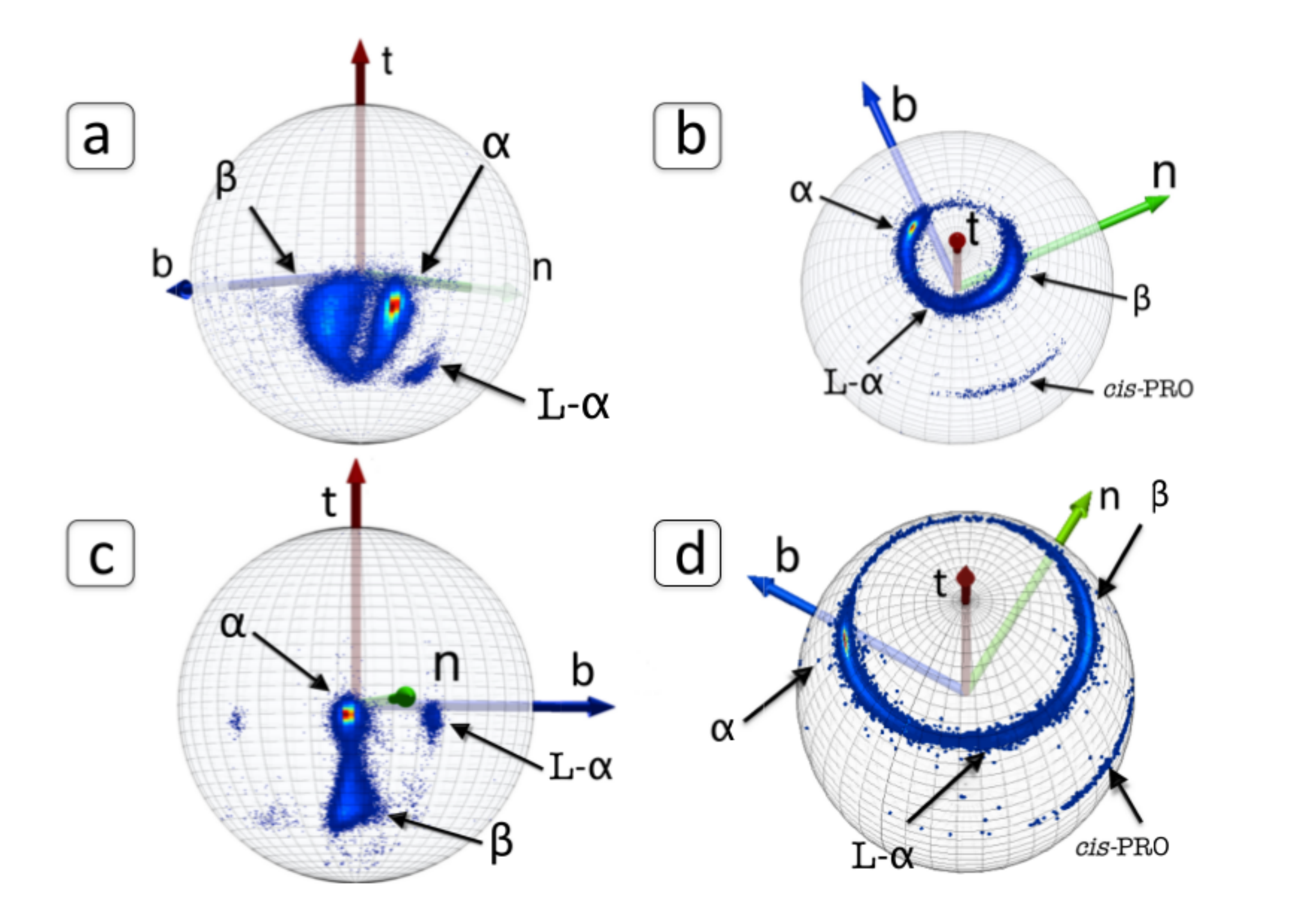}
        \caption{{ 
     (Color online) a) Distribution of C$_\beta$ atoms in the C$_\alpha$ centered Frenet frames in PDB structures that have 
     been measured
with better than 1.0 \AA~resolution. The three major structures $\alpha$-helices, $\beta$-strands and left-handed $\alpha$-
helices have been marked, following their  identification in PDB.  b) Same as a) but for backbone C atoms. Note that C 
atoms that precede a $cis$-proline are clearly identifiable. c) 
Same as a) and b) but for backbone N atoms.  d) Same as a), b) and c) but
for backbone O atoms. As in b) the atoms preceding a $cis$-proline are clearly identifiable. 
       }}
       \label{fig5}
\end{figure}
%
%
%
%
%%%%%%%%%%%%%%%%%%%%%%%%%%%%%%%%%%%%%%%%%%%%%
%
%
we show the ensuing density distributions, on the surface of the C$_\alpha$ centered sphere.
These figures are constructed from
all  the PDB entries that have been measured using diffraction data with better than 1.0 \AA~resolution.

We note clear rotamer structures: The C$_\beta$, C, N and O atoms are each 
localized, and in a manner that depends on the underlying
secondary structure  \cite{Lundgren-2012b}. Both in the case of C$_\beta$ and
N,  the left-handed $\alpha$ region (L-$\alpha$) is a distinct rotamer which is detached
from the rest. In the case of C and O, the 
L-$\alpha$ region is more connected with the other regions. But for C and O,
the region for residues before {\it cis-}prolines becomes detached from the rest.
In the case of  C and C$_\beta$ we do not 
observe any similar isolated and localized  {\it cis-}proline rotamer. 

The C and O rotamers concentrate on a 
circular region, with 
essentially constant latitude angle with respect to the Frenet frame
tangent vector; for the O distribution, the latitude  is larger. The N rotamers form a narrow strip 
in the longitudinal direction, while the map for C$_\beta$ rotamers form a shape that
resembles a horse shoe.

For comparison, in figure \ref{fig6}
%\marginpar{Fig. \ref{fig6}}
%%%%%%%%%%%%%%%%%%%%%%%%%%%%%%%%%%%%%%%%%%%%%
%
%
%
%
%
%figure -6
\begin{figure}[h]
        \centering
                \includegraphics[width=0.45\textwidth]{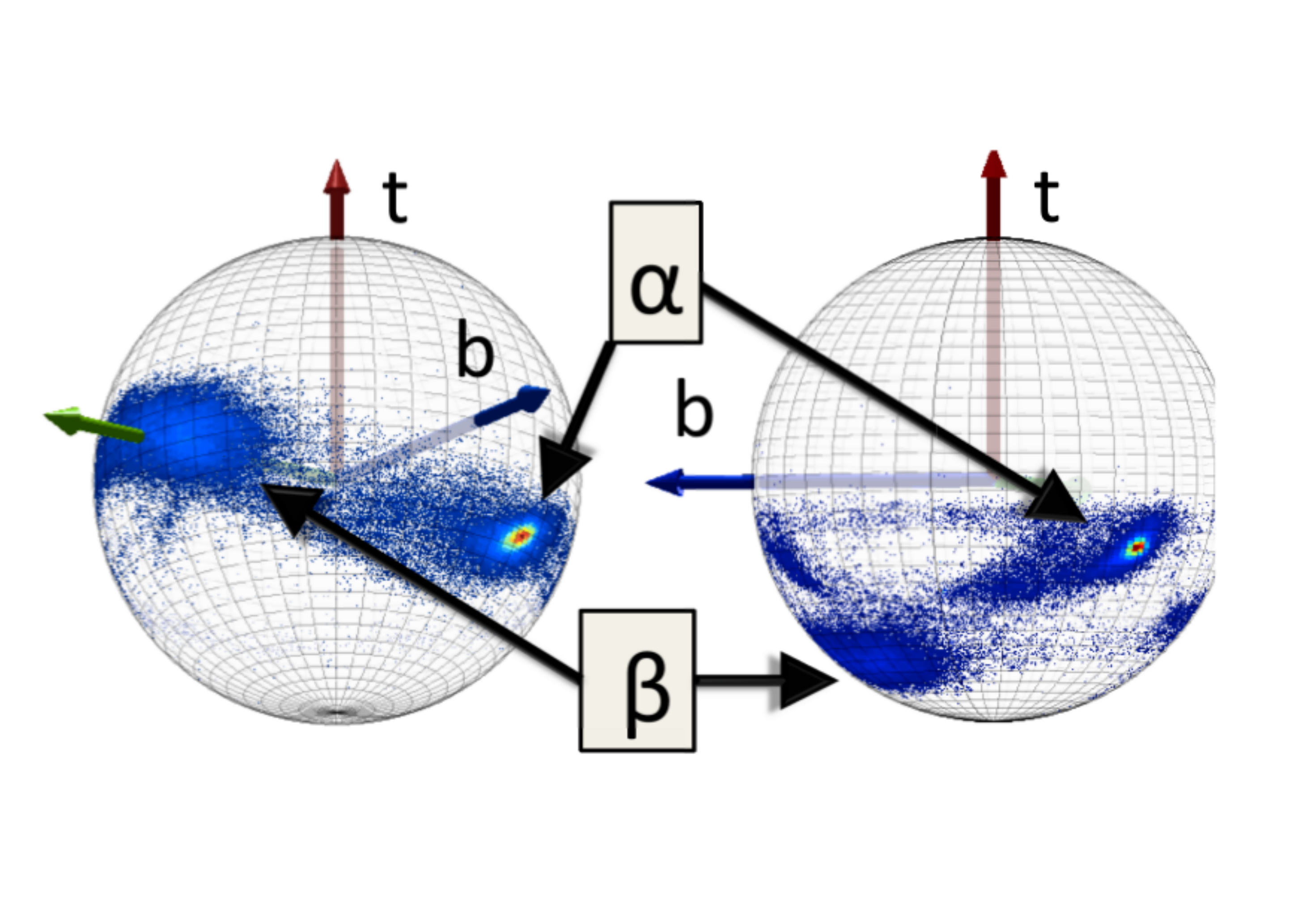}
        \caption{{ 
     (Color online) Distribution of C$_\beta$ atoms (left) and backbone N atoms (right) in the frames of REMO
\cite{Li-2009}.
       }}
       \label{fig6}
\end{figure}
%
%
%
%
%%%%%%%%%%%%%%%%%%%%%%%%%%%%%%%%%%%%%%%%%%%%%
%
%
we visualize  the C$_\beta$ and N distributions in the coordinate system that is utilized in REMO \cite{Li-2009}. 
The secondary structures can be identified, but the rotamers are clearly more delocalized 
than in the case of the Frenet frame map, shown in figure \ref{fig5} a) and c). This delocalization persists in the
case of backbone C and O atoms (not shown). Similarly, we have found that in the case of 
the coordinate system of PULCHRA \cite{Rotkiewicz-2008},  the rotamers are similarly clearly more 
delocalized than in the Frenet frames (not shown). 

One may argue that the stronger the localization of rotamers, the more precise 
will structure analysis, prediction and validation become. From this perspective, 
the  Frenet frames have an advantage over
the frames used {\it e.g.} in PULCHRA and REMO.

The N, C and C$_\beta$ atoms form the covalently bonded heavy-atom corners of the
C$_\alpha$ centered $sp3$-hybridized   tetrahedron. We consider 
the three bond angles  
\begin{eqnarray}
\vartheta_{\rm NC} ~ \simeq ~ {\rm N-C}_\alpha-{\rm C} ~ 
\label{angles1}\\
\vartheta_{{\rm N}\beta} ~ \simeq  ~ {\rm  N-C}_\alpha-{\rm C}_\beta 
\label{angles2}\\
\vartheta_{\beta \rm C} ~ \simeq  ~ {\rm C}_\beta-{\rm C}_\alpha-{\rm C}
\label{angles3}
\end{eqnarray}
The $\vartheta_{\rm NC}$ angle 
relates to the backbone only, while the definition of the other two involves the side chain C$_\beta$. 
In figure \ref{fig7} 
%\marginpar{Fig. \ref{fig7}} 
%%%%%%%%%%%%%%%%%%%%%%%%%%%%%%%%%%%%%%%%%%%%%
%
%
%
%
%
%figure -7
\begin{figure}[h]
        \centering
                \includegraphics[width=0.45\textwidth]{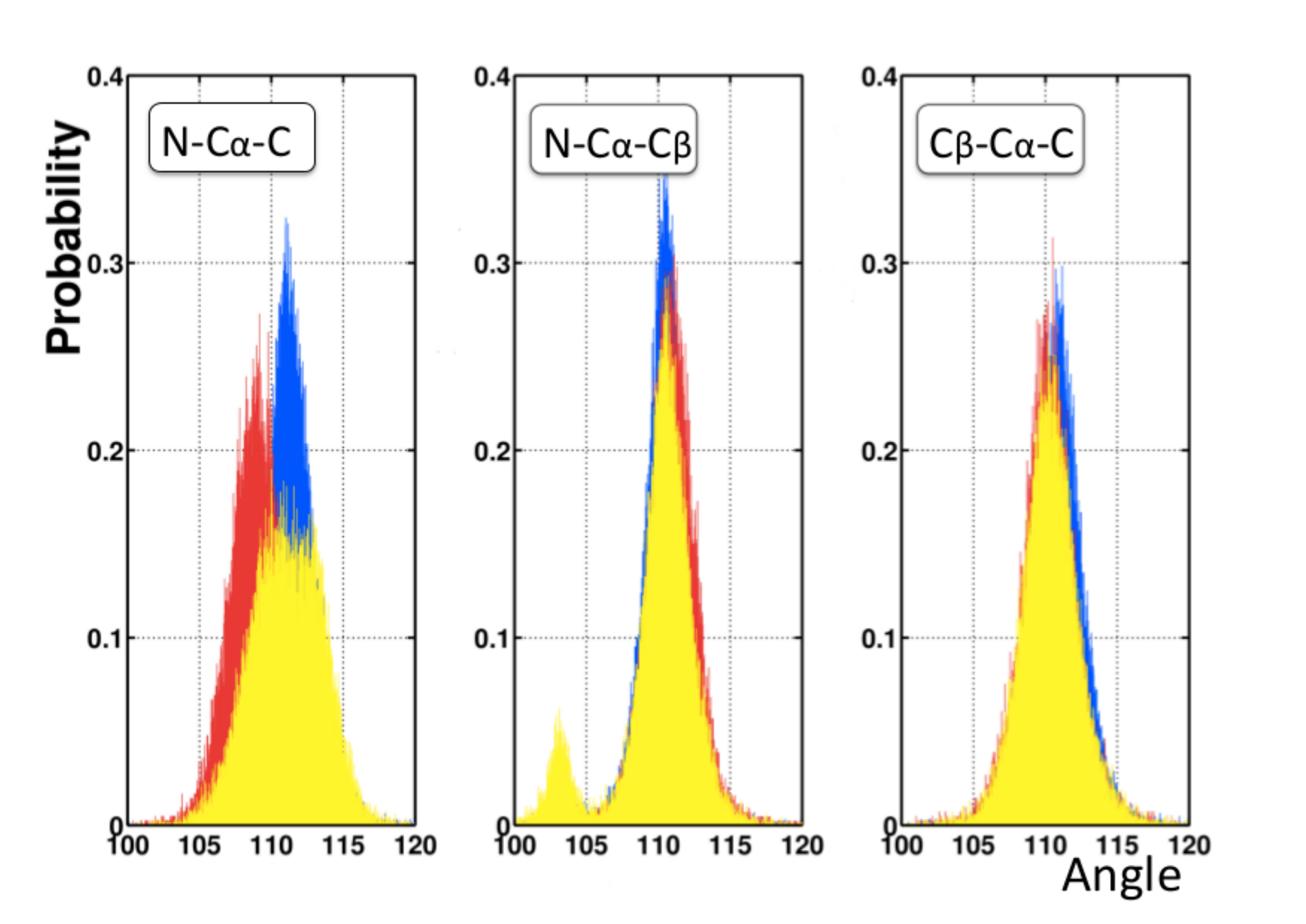}
        \caption{{ 
     (Color online) Distribution of the three bond angles  (\ref{angles1})-(\ref{angles3}), according
to secondary structures.  Blue are $\alpha$-helices, red are $\beta$-strands and yellow are  loops; the small
 (yellow) peak in  N-C$_\alpha$-C$_\beta$ with angle around 103$^{\rm o}$ is due to prolines. See Table 1 for
 the average values for $\alpha$-helices, $\beta$-strands and loops in figure a). See also Table 2 for the average
 values in figures a), b) and c) with no regard to secondary structure. Finally, see Table 3 for the average
 values.
       }}
       \label{fig7}
\end{figure}
%
%
%
%
%%%%%%%%%%%%%%%%%%%%%%%%%%%%%%%%%%%%%%%%%%%%%
%
%
we show the distribution of the three 
tetrahedral bond angles (\ref{angles1})-(\ref{angles3}) in our PDB data set. We find that in 
the case of the two 
side chain C$_\beta$ related angles $\vartheta_{{\rm N}\beta}$ and $\vartheta_{\beta \rm C}$,  
the distribution has a single peak which is compatible with ideal values; 
the isolated small peak in  figure \ref{fig7} b) is due  to $cis$-prolines. 
But in the case of the backbone-only specific angle $\vartheta_{\rm NC}$
we find that in our data set  
this is  not the case. The PDB data set we  use and display  in figure \ref{fig7} a) shows, that there is a 
correlation between the $\vartheta_{\rm NC}$ distribution and the 
backbone secondary structure. See also Table 1.
%
%
%
%
%
%
%%%%%%%%%%%%%%%%%%%%%%%%%%%%%%%%%%%%%%%%
%
%         Table 1
%
%%%%%%%%%%%%%%%%%%%%%%%%%%%%%%%%%%%%%
%
%
%
{
\begin{table}[htb]
  \centering
 \begin{tabular}{lc}      % Alignment for each cell: l=left, c=center, r=right
Structure  & $\vartheta_{{\rm NC}}$     \\
\hline
%GLY     & 113.1 ~ $\pm$ 2.5       & 113.1 ~ $\pm$ 3.4      \\
Helix  &  111.5 ~$\pm$ 1.7   \\
Strand &  109.1 ~$\pm$ 2.0   \\
Loop &  111.0 ~$\pm$ 2.5  \\
\end{tabular}
 \caption{
Average values of the angle $\vartheta_{\rm NC}$ separately  for $\alpha$-helices,
$\beta$-strands and loops in figure \ref{fig7} a) with one-$\sigma$ standard deviations.
}
%\vskip 1.0cm
\end{table}
}

We note that in protein structure validation all three angles (\ref{angles1})-(\ref{angles3}) are commonly 
presumed to assume the ideal values, shown in Table 3.
%
%
%
%
%
%
%%%%%%%%%%%%%%%%%%%%%%%%%%%%%%%%%%%%%%%%
%
%         Table 2
%
%%%%%%%%%%%%%%%%%%%%%%%%%%%%%%%%%%%%%
%
%
%
\begin{table}
\begin{tabular}{llcc}      % Alignment for each cell: l=left, c=center, r=right
Angle   & $\vartheta_{{\rm NC}}$  & $\vartheta_{{\rm C}\beta}$  &   $\vartheta_{\beta{\rm N}}$   \\
\hline
%GLY     & 113.1 ~ $\pm$ 2.5       & 113.1 ~ $\pm$ 3.4      \\
All  &  110.7 ~$\pm$ 2.3 & 110.5 ~$\pm$ 2.0  & 110.3 ~$\pm$ 2.4    \\
PRO & 112.6 ~$\pm$ 2.2 & 111.3 ~$\pm$ 1.7  & 103.2 ~$\pm$ 1.1    \\
rest  & 110.6 ~$\pm$ 2.3  & 110.4 ~$\pm$ 2.0  & 110.7 ~$\pm$ 1.7  \\
\end{tabular}
\caption{Average values of the angles in figures \ref{fig7} computed from our PDB data set, without
subdivision according to secondary structure, and with one-$\sigma$ standard deviations. See also Table 3.}
\end{table}
%
%
%
%
%
%
%%%%%%%%%%%%%%%%%%%%%%%%%%%%%%%%%%%%%%%%
%
%         Table 3
%
%%%%%%%%%%%%%%%%%%%%%%%%%%%%%%%%%%%%%
%
%
%
\begin{table}
\begin{tabular}{llccc}      % Alignment for each cell: l=left, c=center, r=right
Residue & EH-1 & EH-2   &   AK &  TV  \\
\hline
%GLY     & 113.1 ~ $\pm$ 2.5       & 113.1 ~ $\pm$ 3.4      \\
$\vartheta_{{\rm NC}}$(PRO)    & & 112.1  $\pm$ 2.6      & & 112.8  $\pm$ 3.0    \\
$\vartheta_{{\rm NC}}$(REST) & 110.5 & 111.0  $\pm$ 2.7  & 110.4 $\pm$ 3.3 & 111.0  $\pm$ 3.0   \\
$\vartheta_{{\rm C}\beta}$ & 110.1 & & 110.1  $\pm$ 2.9  & \\
$\vartheta_{\beta{\rm N}}$ & 111.2 & & 110.1  $\pm$ 2.8 &  \\
\end{tabular}
\caption{Some average values of the angles in  figure \ref{fig7} reported by various authors, 
together with their one-$\sigma$ standard deviations.}
\end{table}

For example, the deviation of the C$_\beta$ atom from its ideal position is 
among the validation criteria in  MolProbity \cite{Chen-2010}, that  uses it to identify potential 
backbone distortions around C$_\alpha$. 
But several authors \cite{Lundgren-2012b}-\cite{Touw}  
have pointed out that certain variation in the values of the 
$\tau_{\rm NC}$ can be expected, and is in fact present  in PDB data. Accordingly, the protein 
backbone geometry does not  obey the single ideal value paradigm. Since this paradigm motivates
the applicability of small molecule libraries such as the Engh and Huber library \cite{Engh-1991}, \cite{Engh-2001}, 
there is a good case to be made in favor of using  the PDB based libraries \cite{Lovell-2000}, \cite{Dunbrack-1993}, 
\cite{Shapovalov-2011} in the case of proteins.

We remind that  $\vartheta_{\rm NC}$ pertains to the two peptide planes that are 
connected by the C$_\alpha$.
The Ramachandran angles ($\phi,\psi$) are the adjacent dihedrals, but unlike $\vartheta_{\rm NC}$ 
they are specific to a single peptide plane; the Ramachandran angles  
describe the twisting of the ensuing peptide plane. If the 
internal structure of the peptide planes is assumed to be rigid, the flexibility in the bond angle  
$\vartheta_{\rm NC}$ remains the only coordinate that 
can contribute  to the bending of the backbone. Consequently a systematic secondary structure dependence,
as displayed  in  figure \ref{fig7},   is to be expected. 
It could be that the lack of any  observable secondary structure dependence in 
$\vartheta_{{\rm N}\beta}$ and $\vartheta_{\beta \rm C}$ suggests that existing validation methods 
distribute all refinement tension on $\vartheta_{\rm NC}$. 
%It is apparent that there is very little  tension on the two side-chain concomitant angles.  

\subsection{C$_\beta$ atoms}

The side chains are connected to the C$_\alpha$ backbone by the covalent bond between
C$_\alpha$ and C$_\beta$. Consequently  the precision, and high level of localization in the
C$_\beta$ map becomes pivotal  for the construction of accurate 
higher level side chain maps.  

\subsubsection{C$_\beta$ at termini:}
We have analyzed those C$_\beta$ atoms that are located in the immediate proximity of the N and 
the C termini in the PDB data. For this, we have considered the first two C$_\beta$ atoms starting
from  the N terminus, and 
the last two C$_\beta$ atoms that are before the C terminus. 
Note that in the data that describes a crystallographic PDB structure, these do not need to 
correspond to  the actual biological termini of the biological protein. In case 
the termini of the biological protein can 
not be crystallized, the PDB data describes the first two residues
after the N terminus {\it reps.} the last two residues prior to the C terminus
that can be crystallized. Here we consider the termini, as they appear in the PDB data.

Recall, that the termini are commonly located on the surface of 
the protein. As such, they are  accessible to solvent and
quite often oppositely charged. It is frequently presumed that the 
termini are unstructured and highly flexible. They are 
normally not given any regular secondary structure assignment in PDB.
%In Figure 2 we show the individual  {\it terminal segment} C$_\beta$ atoms in our 
%subset of PDB entries with better than 1.0 \AA~resolution. 
But the figure \ref{fig8}  shows that in the C$_\alpha$ Frenet frames the orientations of the two terminal 
C$_\beta$ atoms
\begin{figure}[h]
        \centering
                \includegraphics[width=0.45\textwidth]{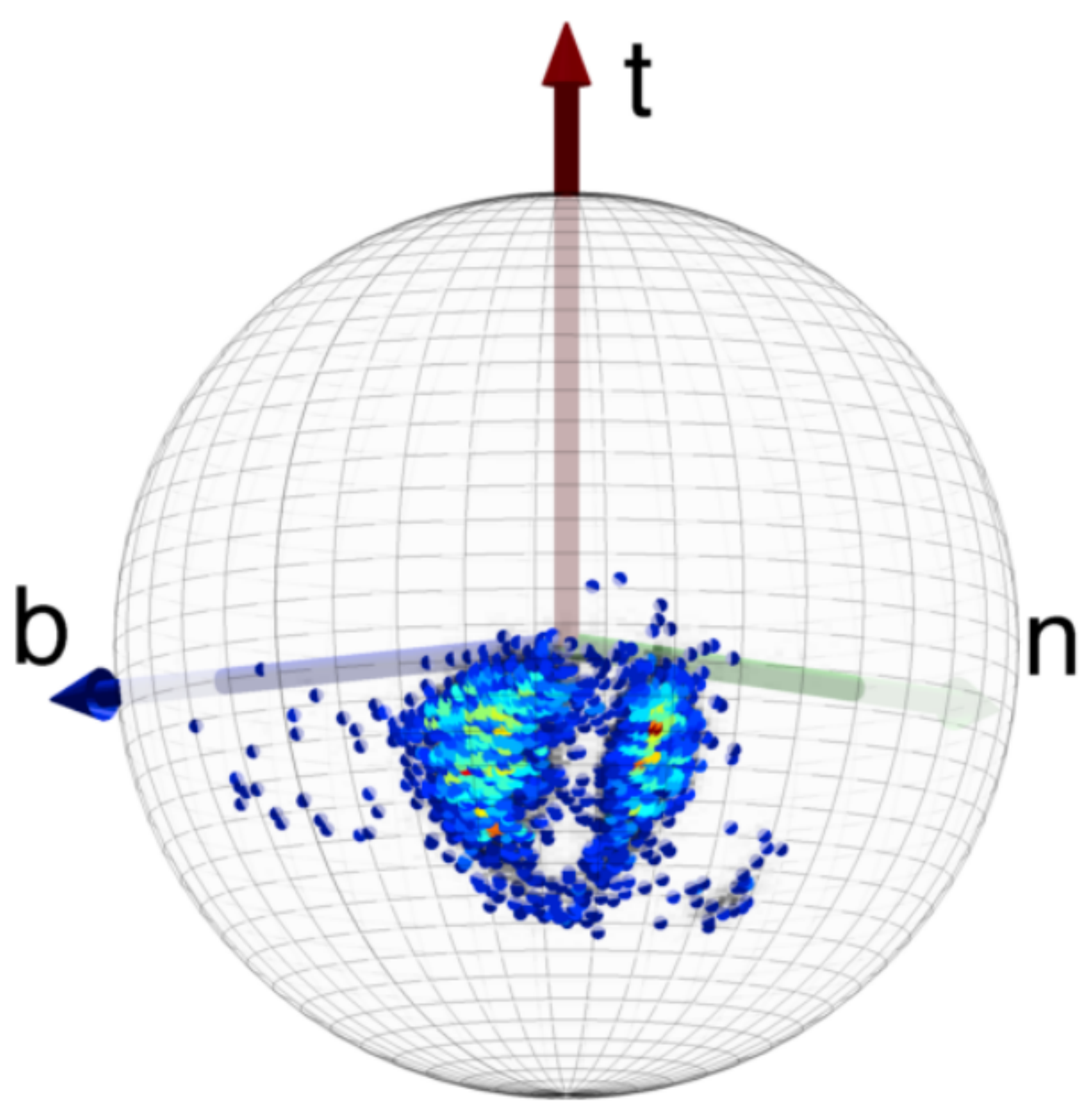}
        \caption{{ 
     (Color online) The distribution of C$_\beta$ directions in the first two and last two residues along PDB structures that 
have been measured using diffraction data with better than 1.0  \AA~resolution. There is no visible difference to
the Figure \ref{fig3} a). In particular, there are very few clear outliers, and they are located  mainly in the region left of the 
main region.        }}
       \label{fig8}
       \end{figure}
are highly regular. Their positions on the surface of the C$_\alpha$ centered sphere 
are fully in  line with that of all the other C$_\beta$ atoms, as shown in figure {\ref{fig5} a). 
In particular,  there are very few outliers. Moreover, the few outliers 
are (mainly) concentrated  in a small region which is located towards the left from the $\beta$-stranded
structures.

%%%%%%%%%%%%%%%%%%%

\subsubsection{C$_\beta$ and proline:}

In figure \ref{fig9}  
\begin{figure}[h]
        \centering
                \includegraphics[width=0.45\textwidth]{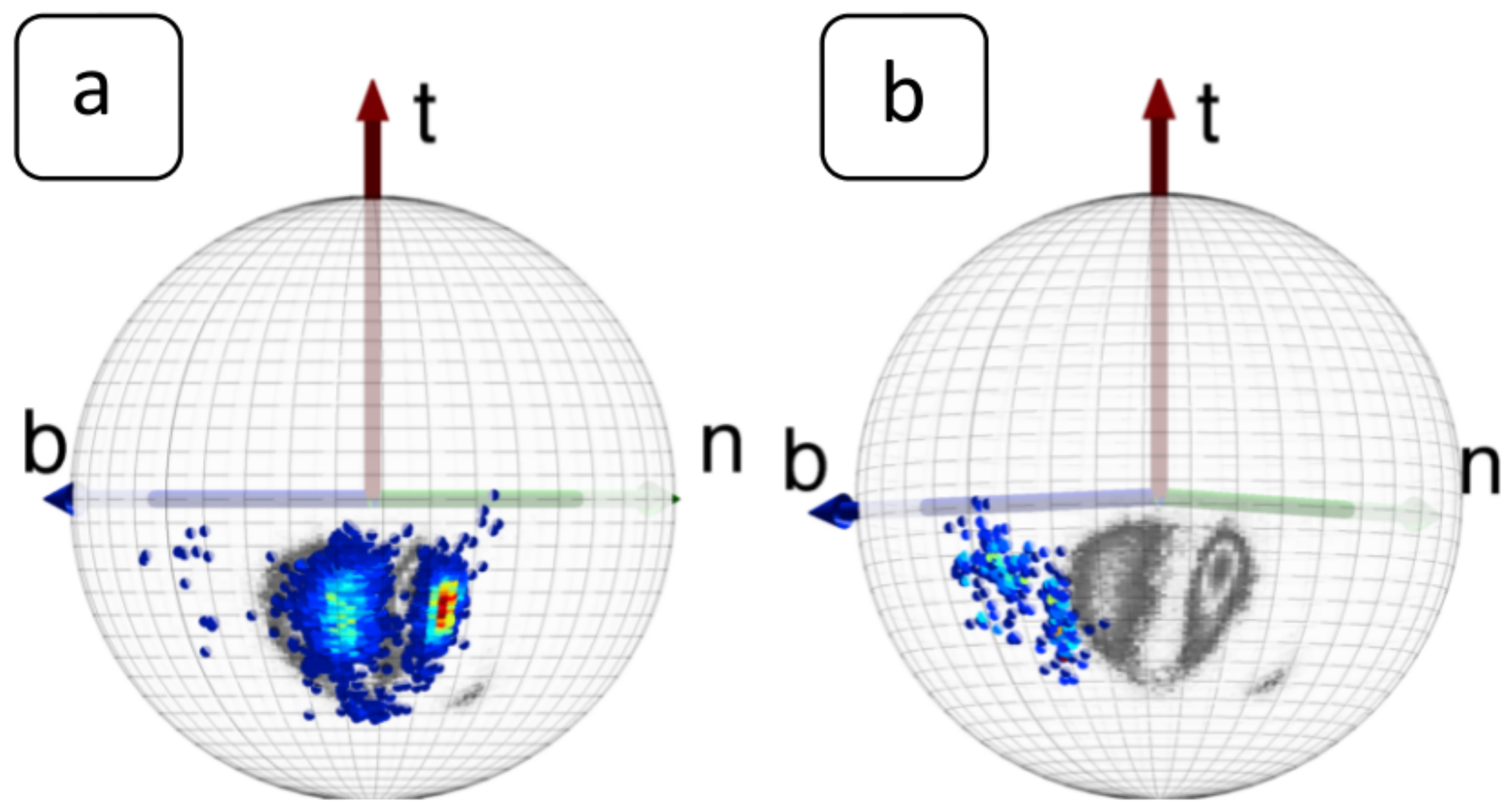}
        \caption{{ 
     (Color online)The distribution of C$_\beta$ in prolines. Figure a) is $trans$-PRO and figure b) is $cis$-PRO.
The grey background is given by Figure \ref{fig5} a).  }}
       \label{fig9}
       \end{figure}
we compare the individual proline contributions in our 
data set with the C$_\beta$ background in figure \ref{fig5} a). 
In figure  \ref{fig9} a) we show the $trans$-proline, and  in figure \ref{fig9} b)  we show the
$cis$-proline. The $trans$-proline has a very good match with the background. There are very few outliers.
These are predominantly located  in the same region as in figure \ref{fig8}, towards the left from the main distribution
{\it i.e.} towards increasing longitude. We observe 
that {\it all} the $cis$-proline are located outside of the main C$_\beta$ 
distribution, towards the increasing longitude from the main distribution.

\vskip 0.2cm
In figures \ref{fig10} a)-d) 
\begin{figure}[h]
        \centering
                \includegraphics[width=0.45\textwidth]{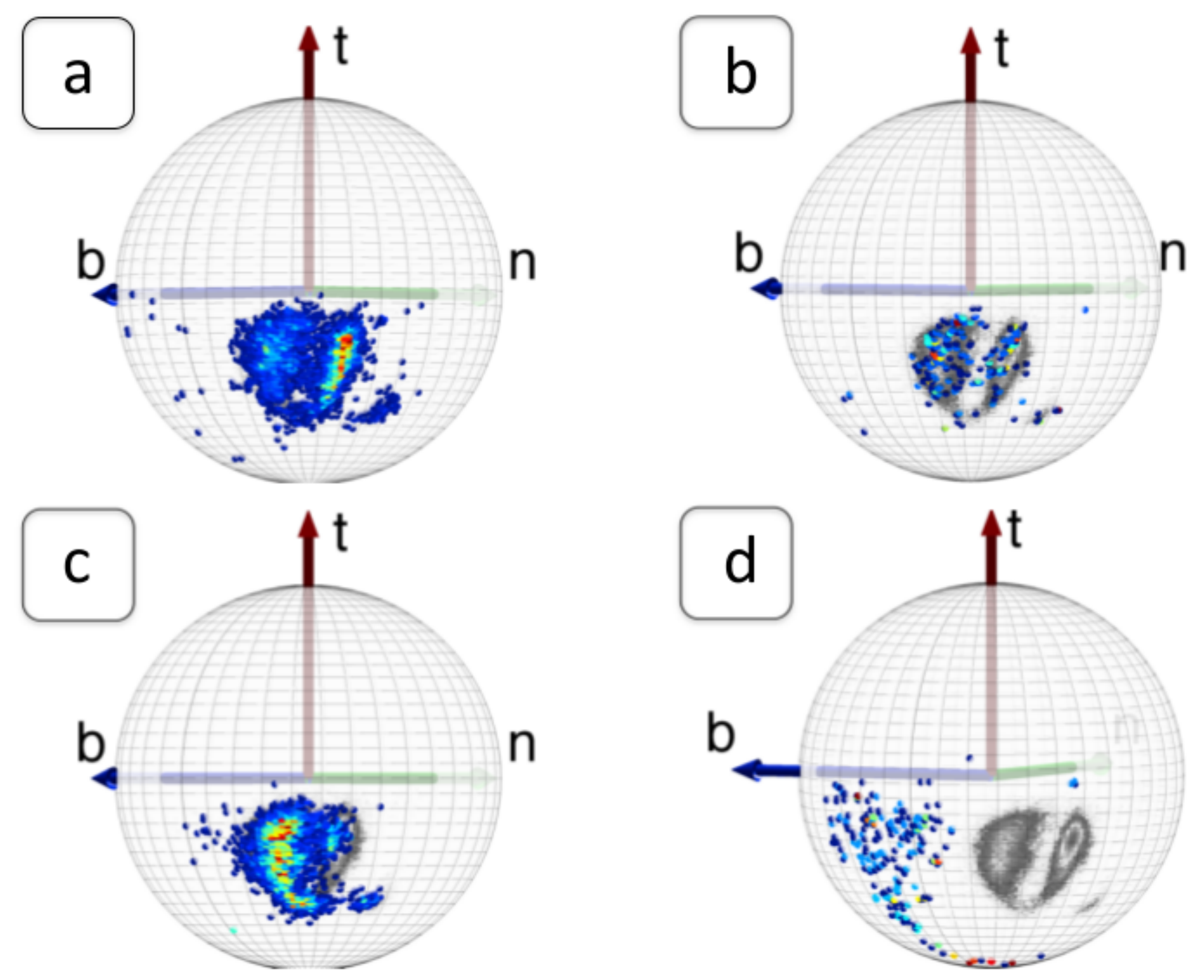}
        \caption{{ 
     (Color online) The distribution of C$_\beta$ atoms immediately after and right before a proline. 
The grey-scaled background is determined by the high-density region of figure \ref{fig5} a). 
In figure a) immediately after  $trans$-PRO and in figure b)
immediately after $cis$-PRO. In figure c) right before $trans$-PRO and in figure d) right before $cis$-PRO. }}
       \label{fig10}
       \end{figure}
we display the C$_\beta$ carbons that are located 
either {\it immediately after}  or {\it right before} 
a proline. We observe the following:

In figure \ref{fig10} a) we have the C$_\beta$ that are immediately
after the $trans$-proline. The distribution matches the background, with very few outliers that are located mostly
in the same region as in figures \ref{fig8}, \ref{fig9} {\it i.e.} towards increasing longitude.  
But there is a {\it very} high density peak in the figure, that overlaps with 
the $\alpha$-helical region: We remind that 
proline is commonly found right before the first residue in a helix.  

In figure \ref{fig10} b) we display those C$_\beta$ atoms which are immediately
after the $cis$-proline. There is again a good match with the background. But unlike in figure \ref{fig10} a)
we also observe a shift towards increasing longitude. in particular, the high density region now coincides
with the $\beta$-stranded region in the background. There are very few outliers, again mainly towards 
increasing longitude.

In figure \ref{fig10} c) we have those C$_\beta$ that are right before a $trans$-proline. 
 There is a clear match with the background distribution. But there are relatively few 
 entries in the $\alpha$-helical position: It is known that 
helices rarely end in a proline.  The 
intensity is very large in the loop region that overlaps the $\beta$-stranded region. 
There are also  a few outliers. Again, the outliers are mainly located in the region towards increasing longitude.

In Figure \ref{fig10} d) we show the C$_\beta$ distribution for residues that are 
right before a $cis$-proline. There are {\it no} entries in
the background region of figure \ref{fig5} a). The distribution is almost 
fully located in the previously observed outlier region, towards the left of the background in the figure.
In addition, we observe an extension of this region towards increasing latitude, reaching all the way to the south-pole. 

\vskip 0.2cm 

Finally, we recall that in figure \ref{fig5} b) 
the region that corresponds to the effect of $cis$-prolines in the preceding C rotamer, is clearly visible. 
But in the case of C$_\beta$ and N atoms, we do not observe any similar 
high density isolated $cis$-region. Consequently the question arises whether the structure of 
the C$_\alpha$ centered covalent tetrahedron is deformed:

In figure \ref{fig11} we show the distribution of the three angles; 
\begin{figure}[h]
        \centering
                \includegraphics[width=0.45\textwidth]{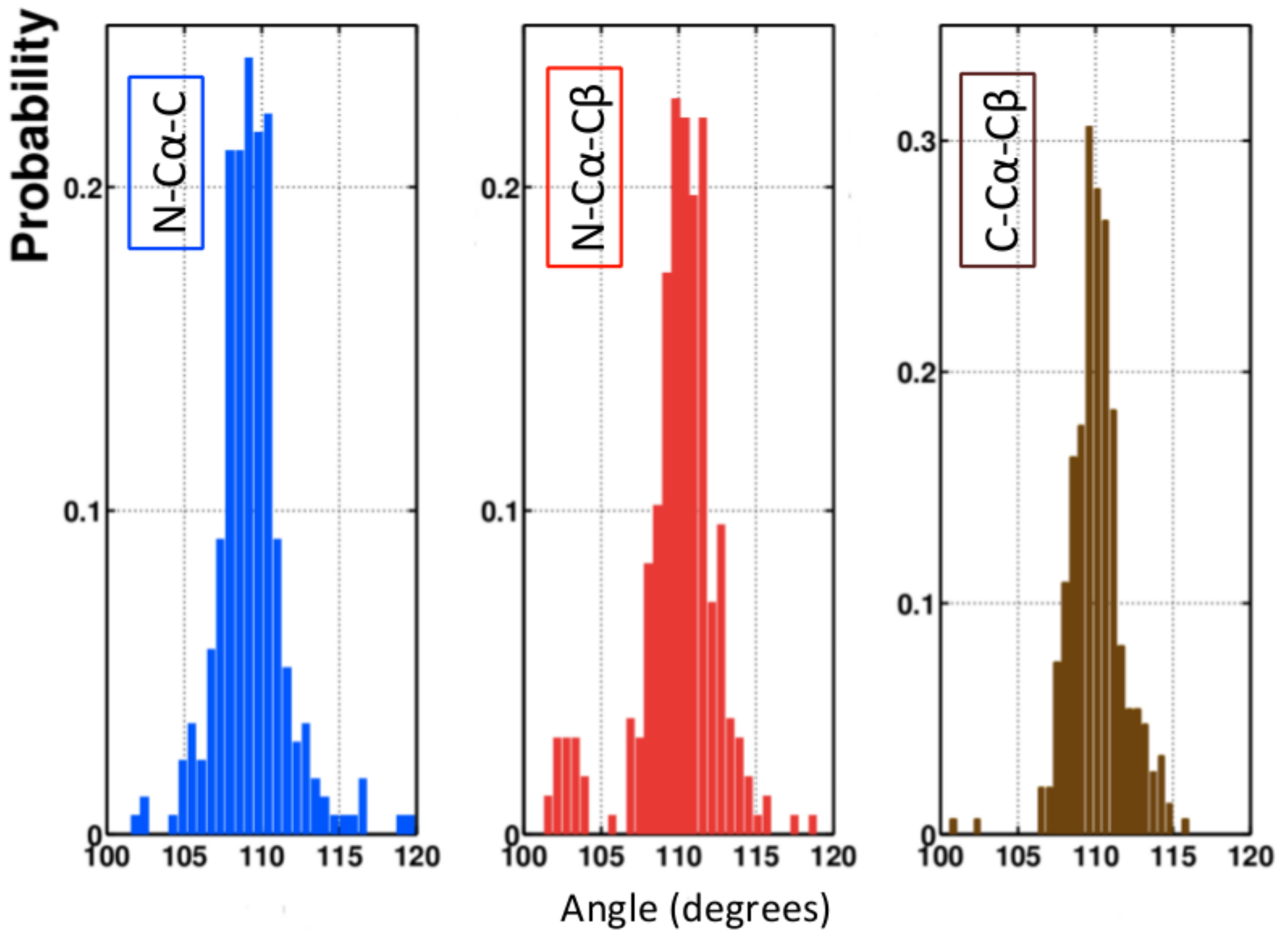}
        \caption{{ 
     (Color online) Distribution of the three heavy atom related angles (in degrees) in the C$_\alpha$ 
centered covalent tetrahedron, in the case of $cis$-proline. The numerical average values together with the
one standard deviations are given in Table 4. }}
       \label{fig11}
       \end{figure}
see also Table 4. 
\begin{table}
\caption{Average values of the angles in figure \ref{fig11}, together with their 
one-$\sigma$ standard deviations.}
\vskip 1.0cm
\begin{tabular}{lccc}      % Alignment for each cell: l=left, c=center, r=right
 Angle    & $\vartheta_{{\rm NC}}$      & $\vartheta_{{\rm C}\beta}$  & $\vartheta_{\beta{\rm N}}$  
 \\
\hline
average     & 109.3 $\pm$ 2.2      & 110.1 $\pm$ 1.8     &  110.0 $\pm$ 2.6    \\
\end{tabular}
\end{table}
We observe a small deviation in the angle N-C$_\alpha$-C. In comparison to proline values 
in Table 2, the value we find in our data set is smaller.

\subsubsection{C$_\beta$ and histidine:}

As another example, in figures \ref{fig12} 
\begin{figure}[h]
        \centering
                \includegraphics[width=0.45\textwidth]{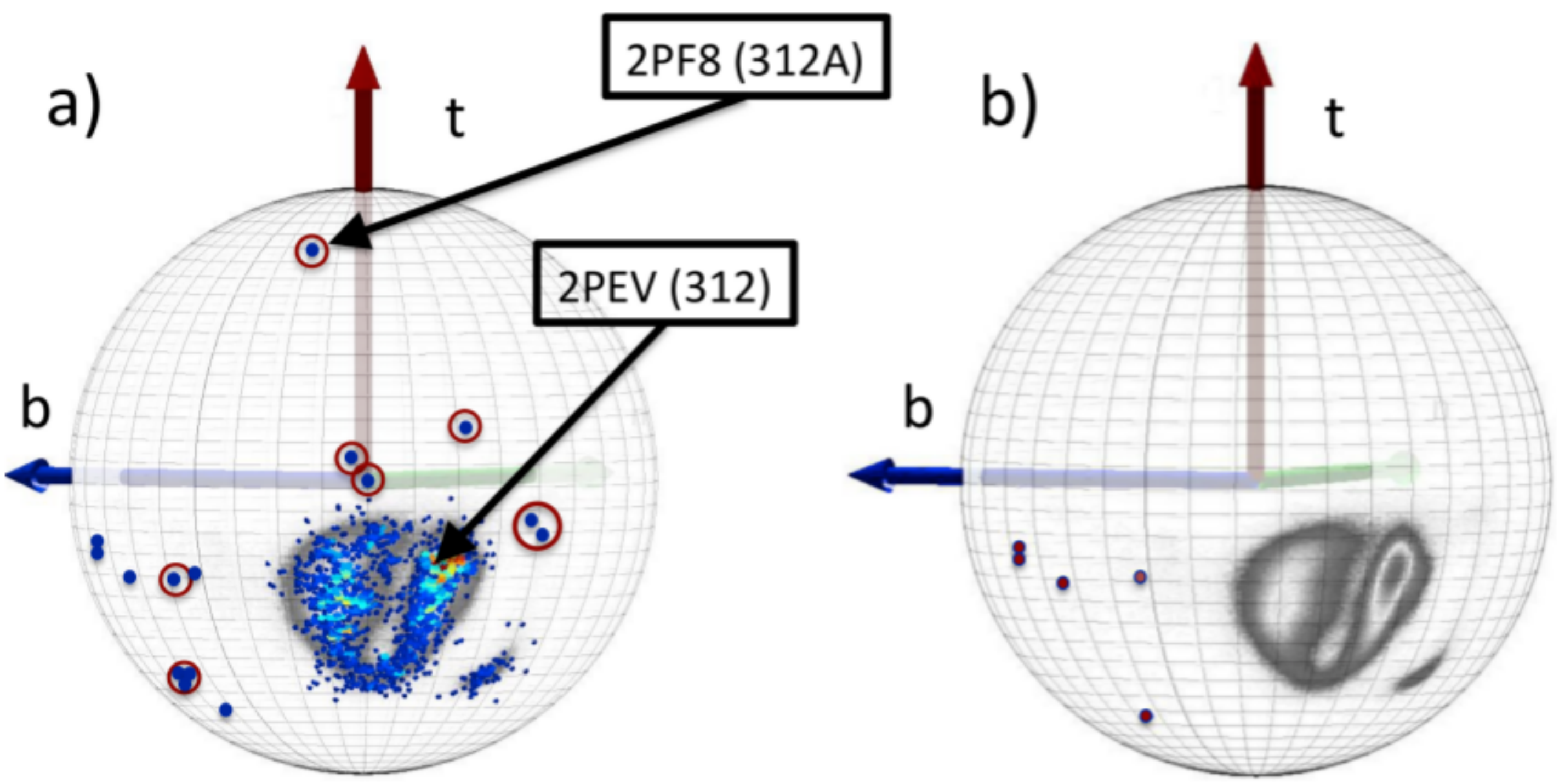}
        \caption{{ 
     (Color online) Figure a) shows C$_\beta$ distribution  of histidine. Some apparent outliers have been encircled, as 
     examples.  
Residue number 
312A in the PDB entry 2PF8 has been identified, together with 312 in the same protein 
2PEV. Figure  b) shows the subset of those HIS that 
precede a $cis$-PRO, there are  five in our data set.}}
       \label{fig12}
       \end{figure}
we display the C$_\beta$ distribution in the case of histidine.  
The figure  \ref{fig12} a)  shows that there is a  very good match with the statistical background distribution. 
There are only a few apparent outliers. Some of them have been encircled, as examples.  
One of the apparent outliers corresponds to the residue number 
312 (HIS) in the PDB entry 2PF8. The latitude is anomalously small. The residue  is located 
relatively close to the C-terminal of the backbone.  
But comparison with figure \ref{fig8}  proposes that this is not the cause for its anomalous latitude
position.
The PDB file of 2PF8 reveals
that this C$_\beta$ atom has two alternative positions. The one we have displayed (312A)
is in an atypical position. The other is not.
This  is also supported by the Frenet frame orientation of the same C$_\beta$
atom 312 in a different
PDB entry of the same protein, with code 2PEV. The C$_\beta$ atom 312 of 2PEV 
is located in the highly populated $\alpha$-helical region.  The reason for the atypical positioning of 312A in 2PF8 
remains to be understood. 

In figure \ref{fig12} b) we plot those HIS that 
precede a $cis$-PRO {\it i.e.} are also present  in Figure \ref{fig10} d).  There are  five 
such entries in HIS. They are all  located in the rotamer that appears to be statistically favored
in figure \ref{fig10} d).
%
%
%
%
%%%%%%%%%%%%%%%%%%%%%%%%%%%%%%%%%%%%%%%%%%%%%%%%%%

%
%
%
%
%%%
%
%
%
%
%
%%%%%%%%%%%%%%%%%%%%%%%%%%%%

%

\subsection{Level-$\gamma$ rotamers}

%%%
%
%
%
%
%
%%%%%%%%%%%%%%%%%%%%%%%%%%%%

\subsubsection{Standard rotamers:} 

We proceed  upwards along the side-chain, to  the 
level-$\gamma$ heavy atoms that are covalently bonded to C$_\beta$.  
Conventionally, these atoms are described by  the side-chain dihedral 
angle $\mathcal X_1$.  This angle is determined by 
the  three covalently bonded heavy atoms C$_\alpha$,  C$_\beta$ and N.
The angle $\mathcal X_1$ determines the dihedral
orientation of the level-$\gamma$ carbon atom, in terms of these three atoms.

We remind that  ALA and GLY do 
not contain any level-$\gamma$ atoms. In the case of ILE and VAL we have two C$_{\gamma}$ while in the case 
of CYS there is a S$_\gamma$ atom.

We first define a $\mathcal X 1$-framing, where the rotamer angle $\mathcal X_1$ appears as a 
dihedral coordinate.  For this we introduce the following C$_\alpha$ based orthonormal triplet  
\begin{equation}
\hskip -5.3cm \mathbf t_{\mathcal X1} \ = \ \frac{ \mathbf r_{\beta} - \mathbf r_{\alpha}  }
{ | \mathbf r_{\beta} - \mathbf r_{\alpha} | }
\label{t1}
\end{equation}
\begin{equation}
\mathbf n_{\mathcal X1} \ = \ \frac{ \mathbf s - \mathbf t_{\mathcal X 1} ( \mathbf s \cdot \mathbf t_{\mathcal X 1} ) }
{| \, \mathbf s - \mathbf t_{\mathcal X 1} ( \mathbf s \cdot \mathbf t_{\mathcal X 1} )\, | } \ \ \ \ \ {\rm where } \ \ \ \ \ 
\mathbf s = \mathbf r_{\alpha} - \mathbf r_{\rm N}
\label{n1}
\end{equation}
\begin{equation}
\hskip -5.5cm \mathbf b_{\mathcal X 1} \ = \ \mathbf t_{\mathcal X1} \times \mathbf n_{\mathcal X1}
\label{b1}
\end{equation}
with $\mathbf r_\alpha$, $\mathbf r_\beta$ and $\mathbf r_N$ 
the coordinates of the pertinent C$_\alpha$, C$_\beta$ and N
atoms, respectively. This constitutes our $\mathcal X 1$-framing, with C$_\alpha$ at the origin. We introduce a sphere
around C$_\alpha$, oriented so that the 
north-pole is in the direction of  $\mathbf t_{\mathcal X1}$. 
Now the dihedral  $\mathcal X_1$ coincides with the ensuing longitude angle.

%
%
%%

%
%
%%%%%%%%%%%%%%%%%%%%%%%%%%%%%%%%%%%%%%%%%%%%%%%%%%

%
%
%
%
%
%
In figures \ref{fig13} we show the distribution of level-$\gamma$ carbon atoms. 
\begin{figure}[h]
        \centering
                \includegraphics[width=0.45\textwidth]{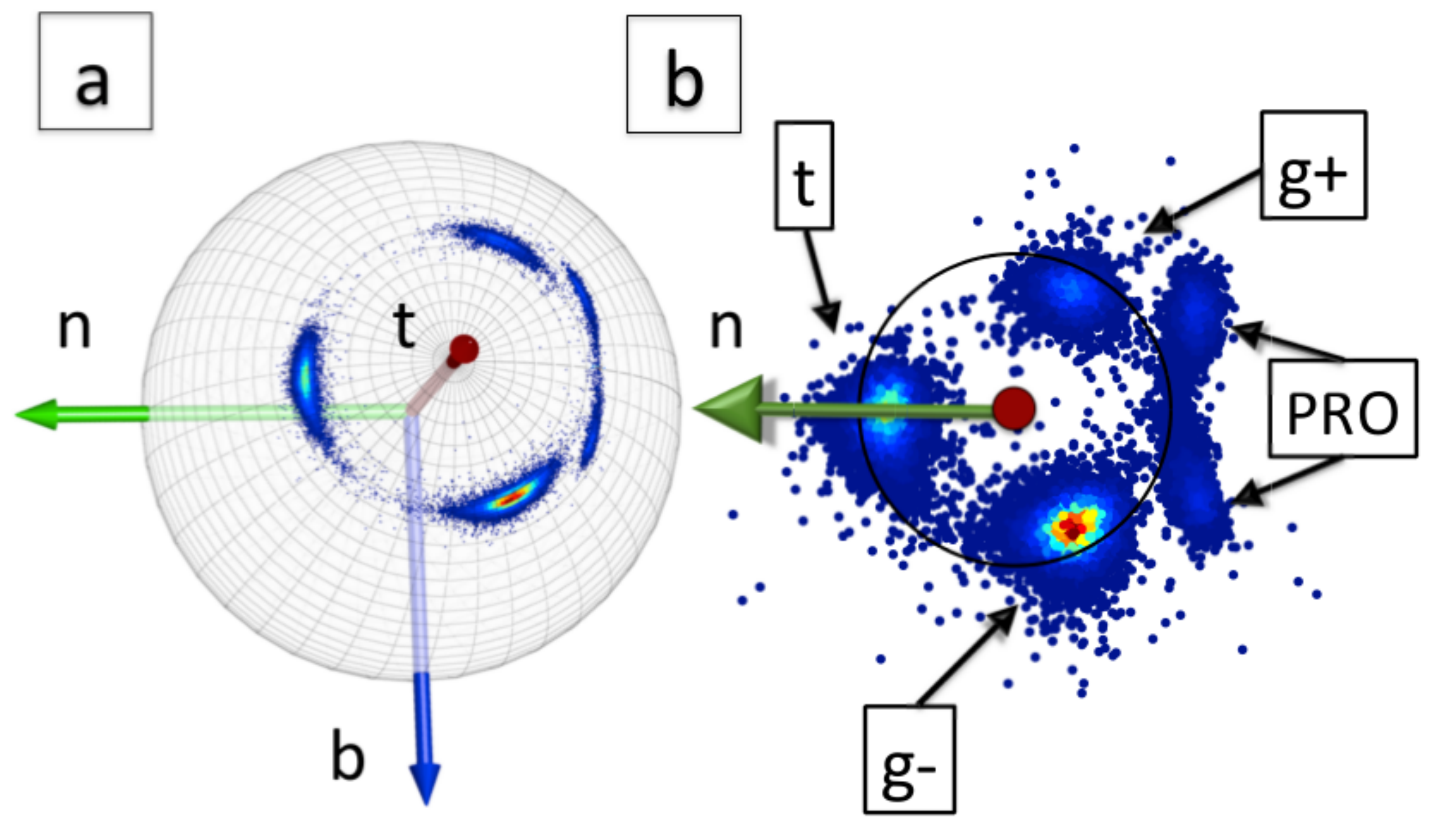}
        \caption{{ 
     (Color online) a) C$_\gamma$ atoms in the $\mathcal X 1$-frames (\ref{t1})-(\ref{b1})
on the C$_\alpha$ centered two-sphere. b) Stereographic projection of a) using (\ref{ste3}).
The three rotamers and proline are identified.}}
       \label{fig13}
       \end{figure}
The figure \ref{fig13} a) shows the distribution on the surface of the 
C$_\alpha$ centered two-sphere.
In figure \ref{fig13} b) we use the stereographic projection (\ref{stereo}) with the choice
\begin{equation}
f(\kappa) = \frac{1}{1+\exp \{\kappa^2\} }
\label{ste3}
\end{equation}
in equation (\ref{ste2}).
The three rotamers {\it gauche$\pm$} (g$\pm$) and {\it trans} (t) have been identified in this figure. 
The prolines are also visible, as rotamers.
In addition, in  figure \ref{fig13} b) we have a circle that
shows the average distance of the data points from the north-pole (origin) 
on the stereographic plane.  A number of apparent outliers are
visible in fig. \ref{fig13} b).  

We note that the underlying secondary structure of the 
backbone is not visible in figures \ref{fig13}. This is a difference 
between figures  \ref{fig5} and \ref{fig13}, in the former the underlying 
backbone secondary structure
is visible in the density profile. 

In figures \ref{fig14} 
\begin{figure}[h]
        \centering
                \includegraphics[width=0.45\textwidth]{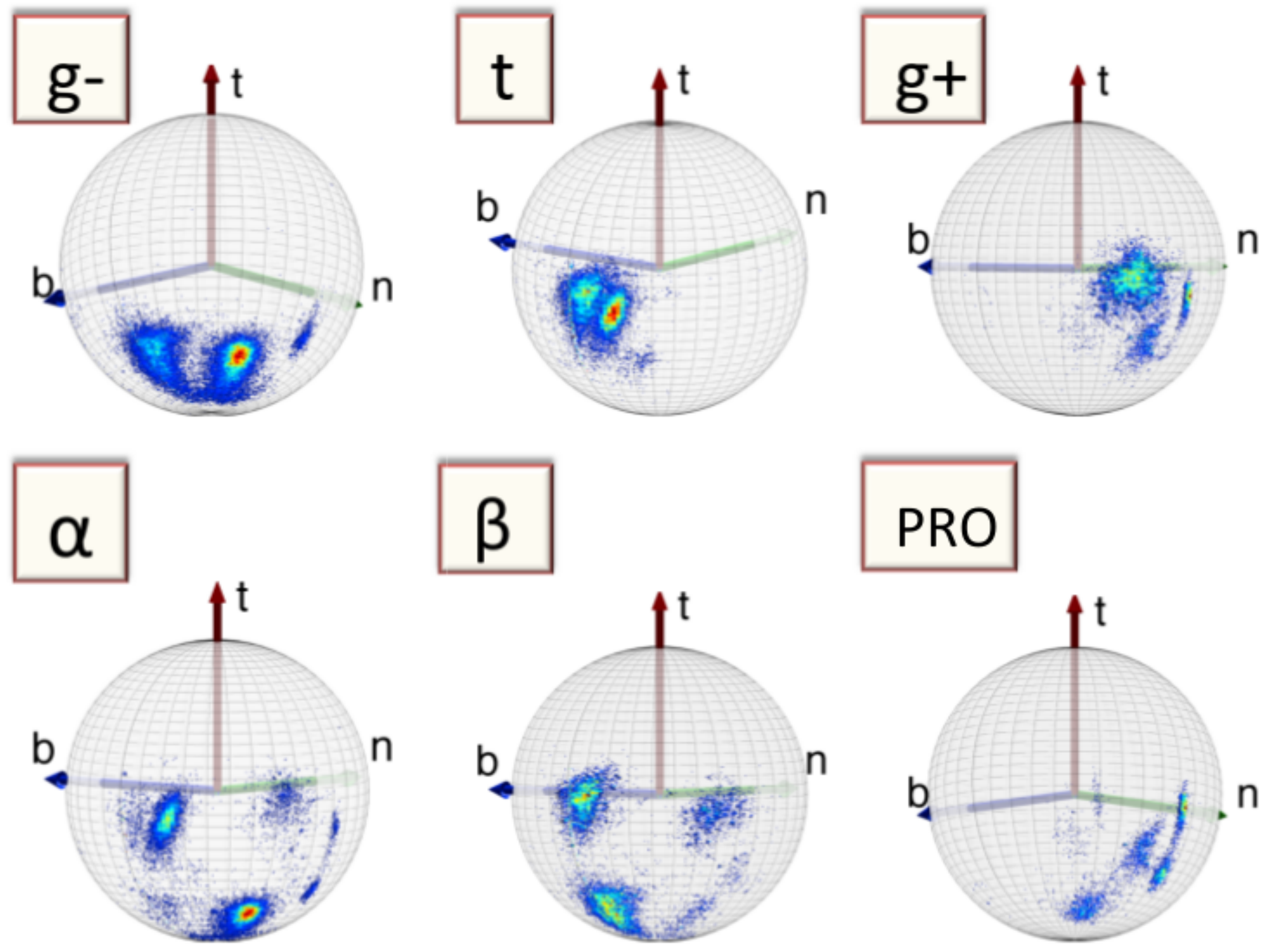}
        \caption{{ 
     (Color online) Frenet frame view of the level-$\gamma$ carbons, separately for the three rotamer states $g\pm$ and $t$ (top line)
and for for $\alpha$-helices, $\beta$-strands and prolines (bottom line). }}
       \label{fig14}
       \end{figure}
we show how 
the C$_\gamma$ atoms are seen by the observer who is located at the C$_\alpha$ atom,  
and oriented according to the backbone Frenet frames; these are the frames
used  in figures \ref{fig5}. \emph{Now} both the rotamer structure and the various  
backbone secondary structures  are clearly seen.

\subsubsection{Secondary structure dependent level-$\gamma$ rotamers:}

In the  C$_\alpha$ Frenet frame figures  \ref{fig14} the secondary structure dependence is visible.
But unlike figure \ref {fig13} a) the  C$_\alpha$  Frenet
frame figures \ref{fig14} lack an apparent symmetry.  This complicates the implementation
of the stereographic projection,  such as the one shown in figure  \ref{fig13} b). 
We proceed to introduce a new set of frames, that enables us to analyze the secondary 
structure dependence of the $\gamma$-level atoms in terms of the stereographic projection: 

We choose the unit length vector $\mathbf t_\beta$, to coincide with
the unit vector that points from C$_\alpha$ at point $\mathbf r_\alpha$ towards  
C$_\beta$ at point $\mathbf r_\beta$. 
\begin{equation}
\mathbf t_\beta = \frac{ \mathbf r_\beta - \mathbf r_\alpha} { |\mathbf r_\beta - \mathbf r_\alpha| } 
\label{tbeta}
\end{equation}
We use the next C$_\alpha$ atom along the backbone, to define the following unit length
vector
 \begin{equation}
\mathbf n_\beta = \frac{\mathbf t_\beta \times \mathbf t_\alpha}{| \mathbf t_\beta \times \mathbf t_\alpha |}
%
%\frac{ \mathbf t^\alpha -  (\mathbf t^\alpha \cdot \mathbf n^\beta) \mathbf n^\beta }
%{ \mathbf t^\alpha -  (\mathbf t^\alpha \cdot \mathbf n^\beta) \mathbf n^\beta }
\label{nbeta}
\end{equation}
Here $\mathbf t_\alpha$ is the vector (\ref{t}).  The orthonormal triplet is completed by
\begin{equation}
\mathbf b_\beta = \mathbf t_\beta
\times \mathbf n_\beta
\label{bbeta}
\end{equation}
We may choose either C$_\alpha$ or C$_\beta$ to coincide with  the origin; the C$_\alpha$ centered coordinate
system is the original roller coasting observer while the C$_\beta$ centered coordinate system corresponds to 
an observer who has climbed "one-step-up" along the side chain. We
map the level-$\gamma$ atoms
on the surface of the pertinent, surrounding two-spheres. In figures \ref{fig15} a) and b) 
\begin{figure}[h]
        \centering
                \includegraphics[width=0.45\textwidth]{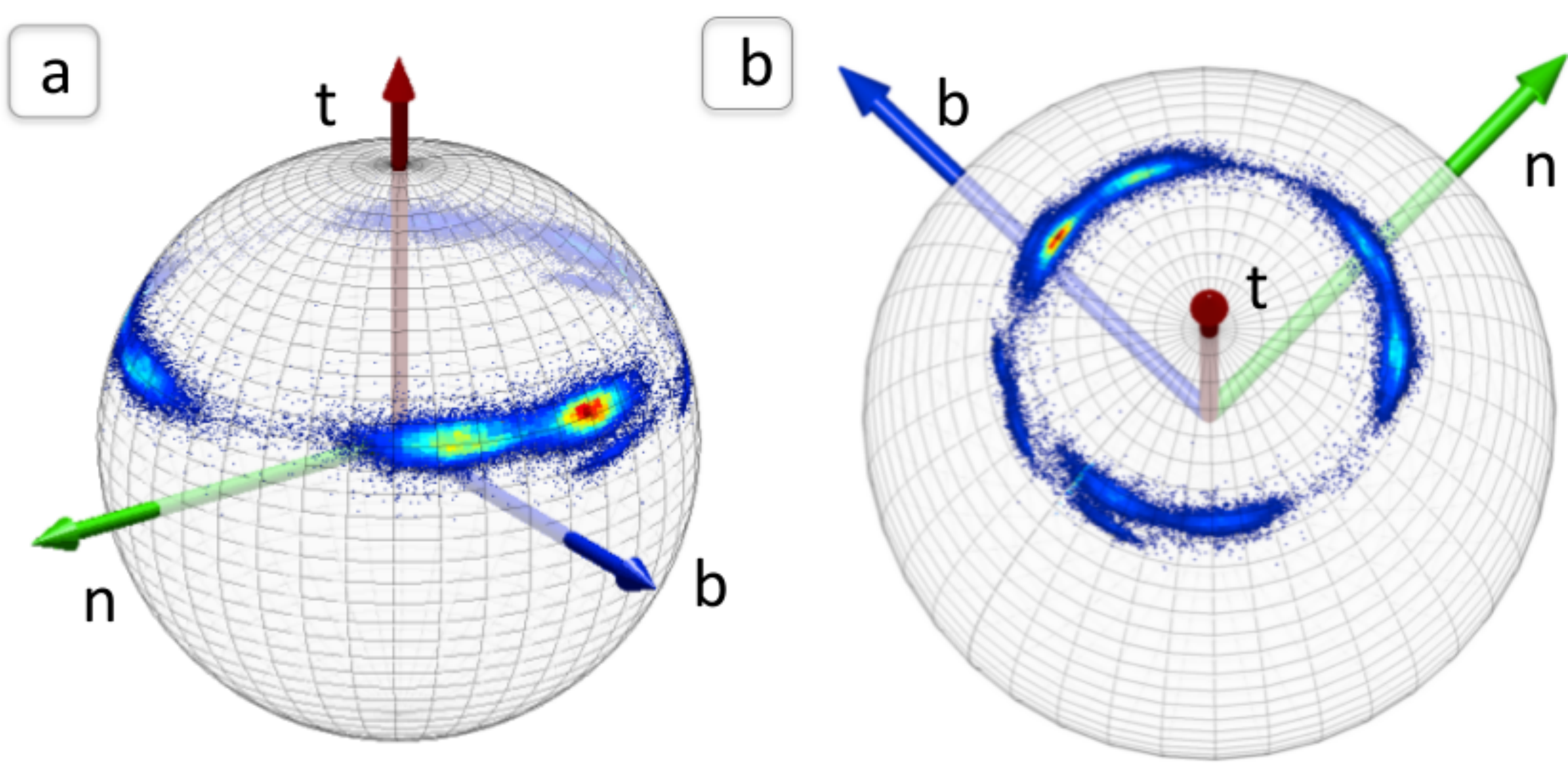}
        \caption{{ 
     (Color online)The level-$\gamma$ atoms as seen in the coordinates (\ref{tbeta})-(\ref{nbeta}). In a) the origin
coincides with the C$_\alpha$ atom, in b) it coincides with the C$_\beta$ atom. }}
       \label{fig15}
       \end{figure}
we show the results.
There is very little qualitative difference between the C$_\alpha$ and  C$_\beta$ centered
distributions, except for latitude {\it i.e.}  the distance from the north-pole. The distributions resemble those in figure \ref{fig13} a),
except that there is additional fine structure: The secondary structures 
are now clearly separated from each other into disparate rotamers.

In figure \ref{fig16} 
\begin{figure}[h]
        \centering
                \includegraphics[width=0.45\textwidth]{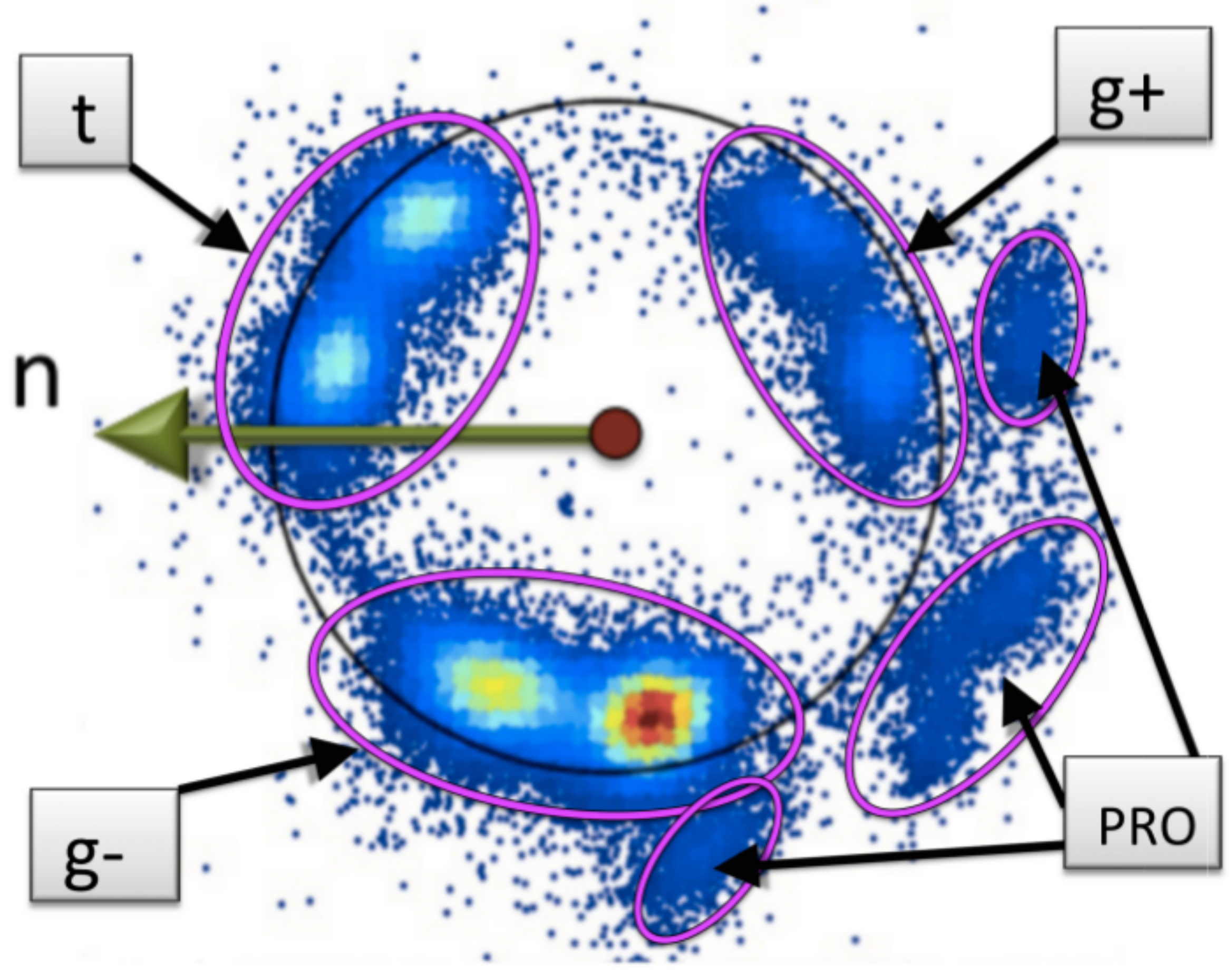}
        \caption{{ 
     (Color online) Stereographic projection of 
level-$ \gamma $ rotamers in the frame of
figure \ref{fig15} b) in combination with (\ref{ste3}). }}
       \label{fig16}
       \end{figure}
 we have stereographic projected   figure \ref{fig15} b), in combination with the map (\ref{ste3}). 
In figures \ref{fig17} we identify the rotamers according to
\begin{figure}[h]
        \centering
                \includegraphics[width=0.45\textwidth]{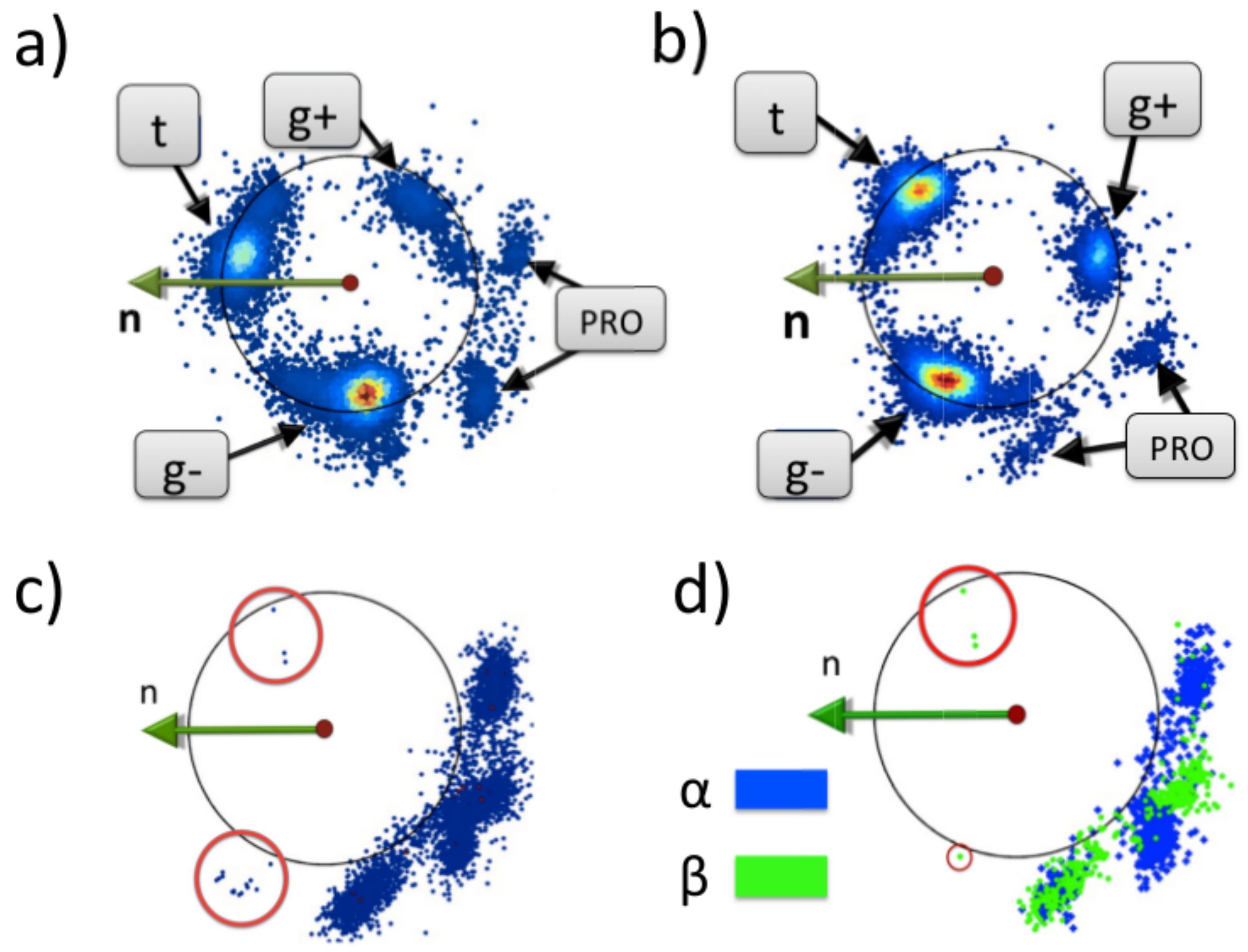}
        \caption{{ 
     (Color online) The identification of rotamers and major secondary structures  in figure \ref{fig16}. In a) the
$\alpha$-helices, in b) the $\beta$-strands. In c) all prolines, and in d) prolines divided according to
their $\alpha$-helical (blue) and $\beta$-stranded (green) assignment in PDB. Some apparent outliers 
have been high-lighted with red circles.}}
       \label{fig17}
       \end{figure}
the $\alpha$-helical and $\beta$-stranded regions, and
the rotamers for prolines. 

The $\alpha$-helical rotamer distribution in  figure \ref{fig17} a) and 
$\beta$-stranded distribution in figure \ref{fig17} b) 
have essentially the same latitude angle. But there
is a visible difference in the longitudes. Each  has a trimodal structure, and  
we again denote the rotamers as g$\pm$ and t.
The distributions are related to each other by 120$^{\rm o}$ longitudinal rotations. 
It is noteworthy how the prolines shown in figures \ref{fig17} c) and d) also  reflect the 
backbone secondary structure, as assigned by PDB.
In these figures  
we have also  highlighted some apparent outlying prolines. These are located in two clusters. 

There are also outliers that are outside of the range of the stereographic projection in figures \ref{fig17}.
The projection - to the extent it has been plotted - 
covers a disk-like region  around the north-pole {\it i.e.}  around the 
tip of vector $\mathbf t$ in  the figure.  The far-away outliers  can be visualized by properly rotating the sphere.
The rotated sphere is shown in figure \ref{fig18}. 
\begin{figure}[h]
        \centering
                \includegraphics[width=0.45\textwidth]{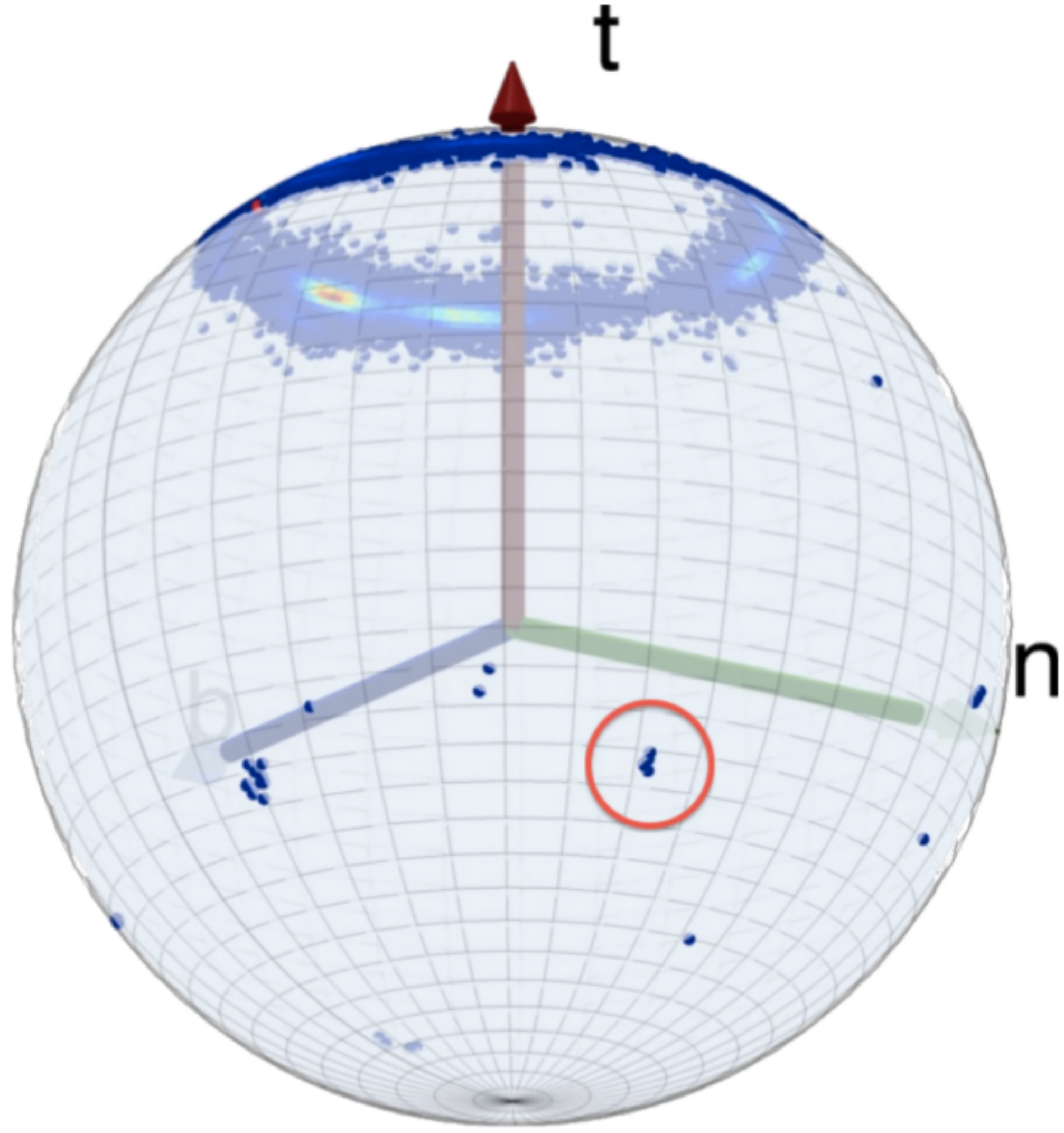}
        \caption{{ 
     (Color online) An example of a group of five 
far away outliers in figure \ref{fig15} b), made visible by rotating the sphere and highlighting with 
red circle. They all correspond to the same 
protein but with different PDB codes:  1FN8, 1FY4, 1FY5,  1GDN and 1GDQ. In each case, the outlier is in the 
residue number 65 A. Note that there are also several other far away outliers.}}
       \label{fig18}
       \end{figure}
A number of far-away outliers  are now visible. 
As an example, we have encircled one group of outliers. It pertains to the mutually related PDB entries
1FN8, 1FY4, 1FY5, 1GDN and 1GDQ. These outliers all have the same residue number 65 in the PDB data. 
It is a  multiple position entry and the figure shows that one of these (A) is atypical.

Finally, as a concrete example of an amino acid we consider threonine,  where the level-$\gamma$ consists of 
a C$_\gamma$ and O$_\gamma$  pair. 
In figure \ref{fig19} a) 
\begin{figure}[h]
        \centering
                \includegraphics[width=0.45\textwidth]{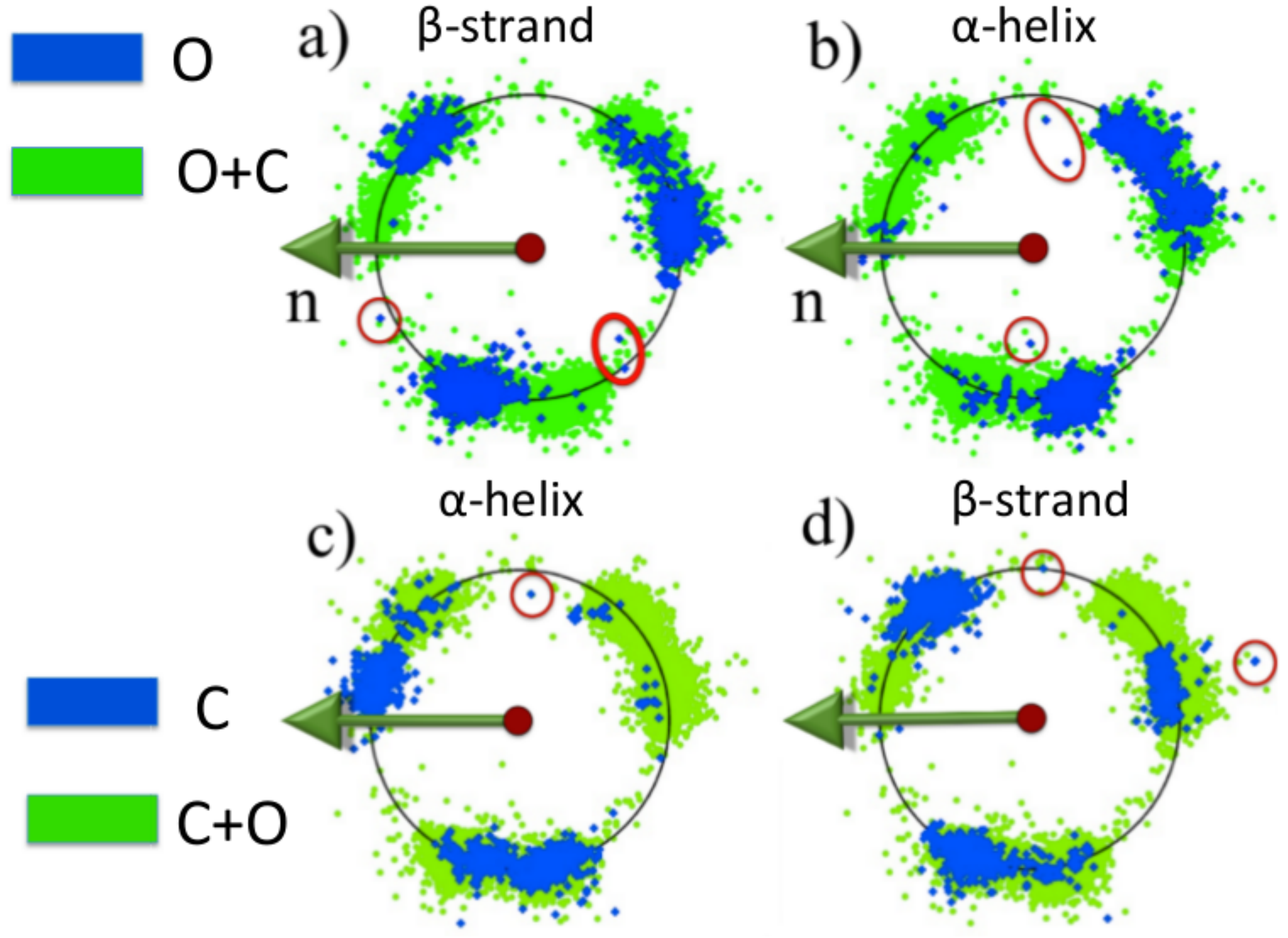}
        \caption{{ 
     (Color online) 
a) O$_\gamma$ (dark blue) in THR, with backbone in the $\beta$-stranded position. 
b) O$_\gamma$ (dark blue) in THR, with backbone in the $\alpha$-helix position.  c) Same as a) but for C$_\gamma$.
d) Same as b) but for C$_\gamma$. Some apparent outliers are encircled.
The (light green)  background  in each Figure consists of all O$_\gamma$ and C$_\gamma$ in THR.}}
       \label{fig19}
       \end{figure}
we display (in blue) those  O$_\gamma$ atoms where the backbone is
in a $\beta$-strand position according to PDB. In \ref{fig19} b)  we have (in blue)
those  O$_\gamma$ where the backbone is
in an $\alpha$-helix  position.  In figures  \ref{fig19} c) and d)  we 
have the corresponding distributions for C$_\gamma$. 
The (green) background is made of all O$_\gamma$ and C$_\gamma$ atoms in our data set.  
Both the trimodal rotamer
structure and its secondary structure dependence are clearly visible, both in O$_\gamma$ and in C$_\gamma$.
For the latter, the distribution matches that displayed in Figures \ref{fig17} a) and b).   Some apparent outliers have
also been highlighted in Figures \ref{fig19} by encircling them (with red).

%%%
%
%
%
%
%
%%%%%%%%%%%%%%%%%%%%%%%%%%%%

\subsection{Level-$\delta$ rotamers}

%%%
%
%
%
%
%
%%%%%%%%%%%%%%%%%%%%%%%%%%%%

\subsubsection{Standard dihedral angle:}

We proceed upwards along the side-chain, to describe level-$\delta$ atoms.  We start 
with a coordinate  frame which is  centered at the C$_\gamma$ atom. We note that in the 
case of ILE, two alternatives exist and we choose the C$_\gamma$ carbon which is 
covalently bonded to the C$_\delta$ atom. 

We set
\[
\mathbf t_{\mathcal X 2} = \frac{ \mathbf r_{\gamma} - \mathbf r_{\beta}} { |\,
\mathbf r_{\gamma} - \mathbf r_{\beta}\, | } 
\]
and we choose
\[
\mathbf n_{\mathcal X 2}  = \frac{ \mathbf t_{\mathcal X 2}  \times \mathbf t_{\mathcal X 1}  }{| \,
 t_{\mathcal X 2}  \times \mathbf t_{\mathcal X 1} \, | }
\]
The third vector $\mathbf b_{\mathcal X 2}$ that completes  the right-handed
orthonormal triplet 
%($\mathbf n^\gamma, \mathbf b^\gamma, \mathbf t^\gamma$)
is given by
\[
\mathbf b_{\mathcal X 2} = \mathbf t_{\mathcal X 2} \times \mathbf n_{\mathcal X 2} 
\]
In figure \ref{fig20}  
\begin{figure}[h]
        \centering
                \includegraphics[width=0.45\textwidth]{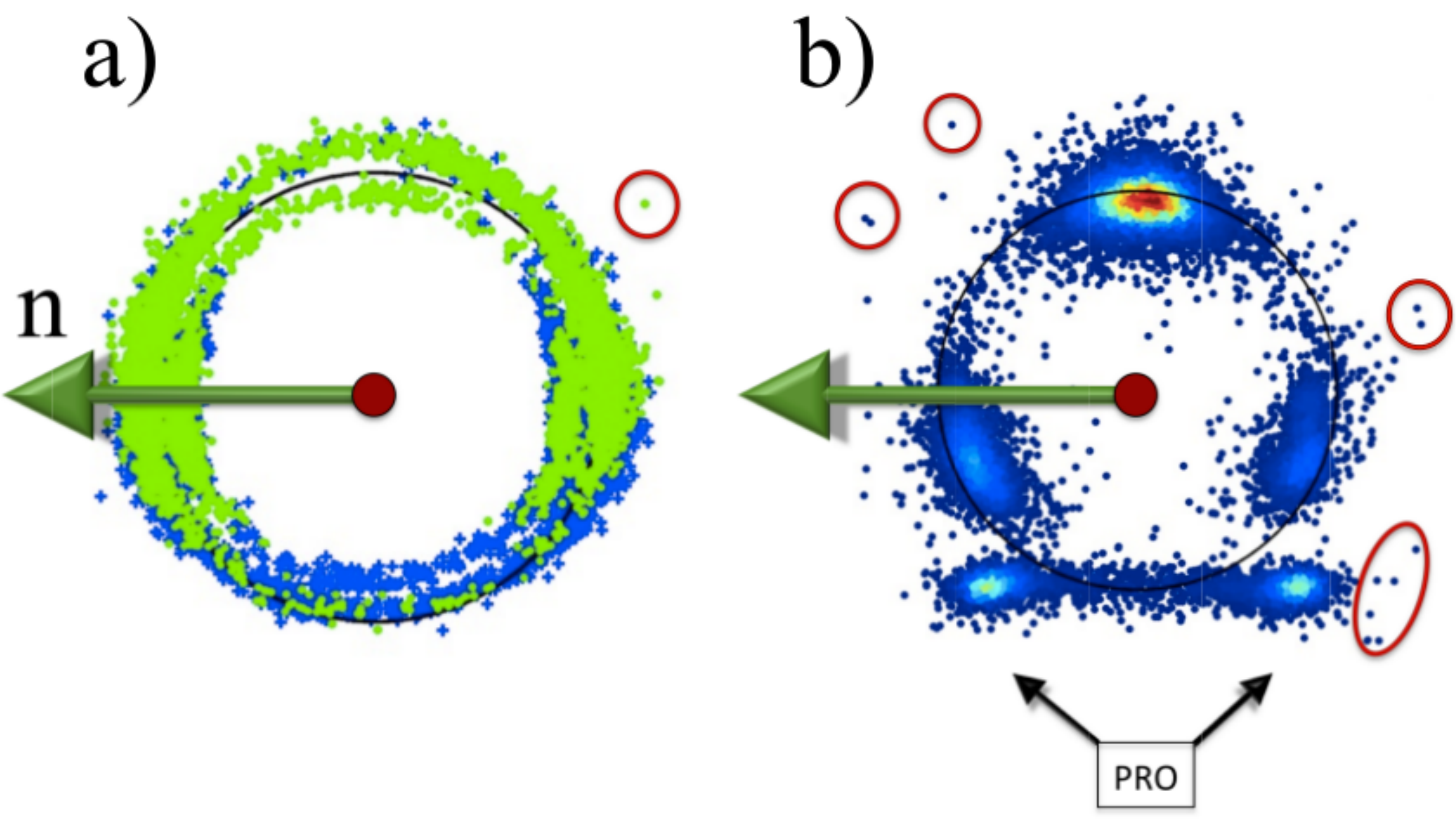}
        \caption{{ 
     (Color online) 
a) Distribution of aromatic and b) non-aromatic  level-$\delta$ C atoms, in the 
stereographic projection  of the unit two-sphere centered at  
the C$\gamma$ atom. In a) the (dark) blue is C$\delta 1$
and (light) green is C$\delta 2$. Some outliers have been encircled, as examples.
The (black) circles
around the center denote the average distance of the distribution.
}}
       \label{fig20}
       \end{figure}
we show the distribution of heavy atoms in level-$\delta$, after  stereographic
projection (\ref{stereo}). The longitude in these figures coincides with the standard $\mathcal X_2$ dihedral
angle, modulo a global $\pi/2$ rotation around the center.
In addition, we introduce the following version of (\ref{ste2})
\begin{equation}
f(\theta) = \frac{1}{1+\theta^4}
\label{ste4}
\end{equation}
In the figure \ref{fig20}, we have separately displayed  the distribution of the
aromatic (a) and the non-aromatic (b) amino acids; we find that starting at level-$\delta$ this  is a 
convenient bisection. A clear trimodal rotamer structure is  present in figure \ref{fig20} b). Some outliers have
been highlighted with circles, as generic examples.
In figure \ref{fig21} a) 
\begin{figure}[h]
        \centering
                \includegraphics[width=0.45\textwidth]{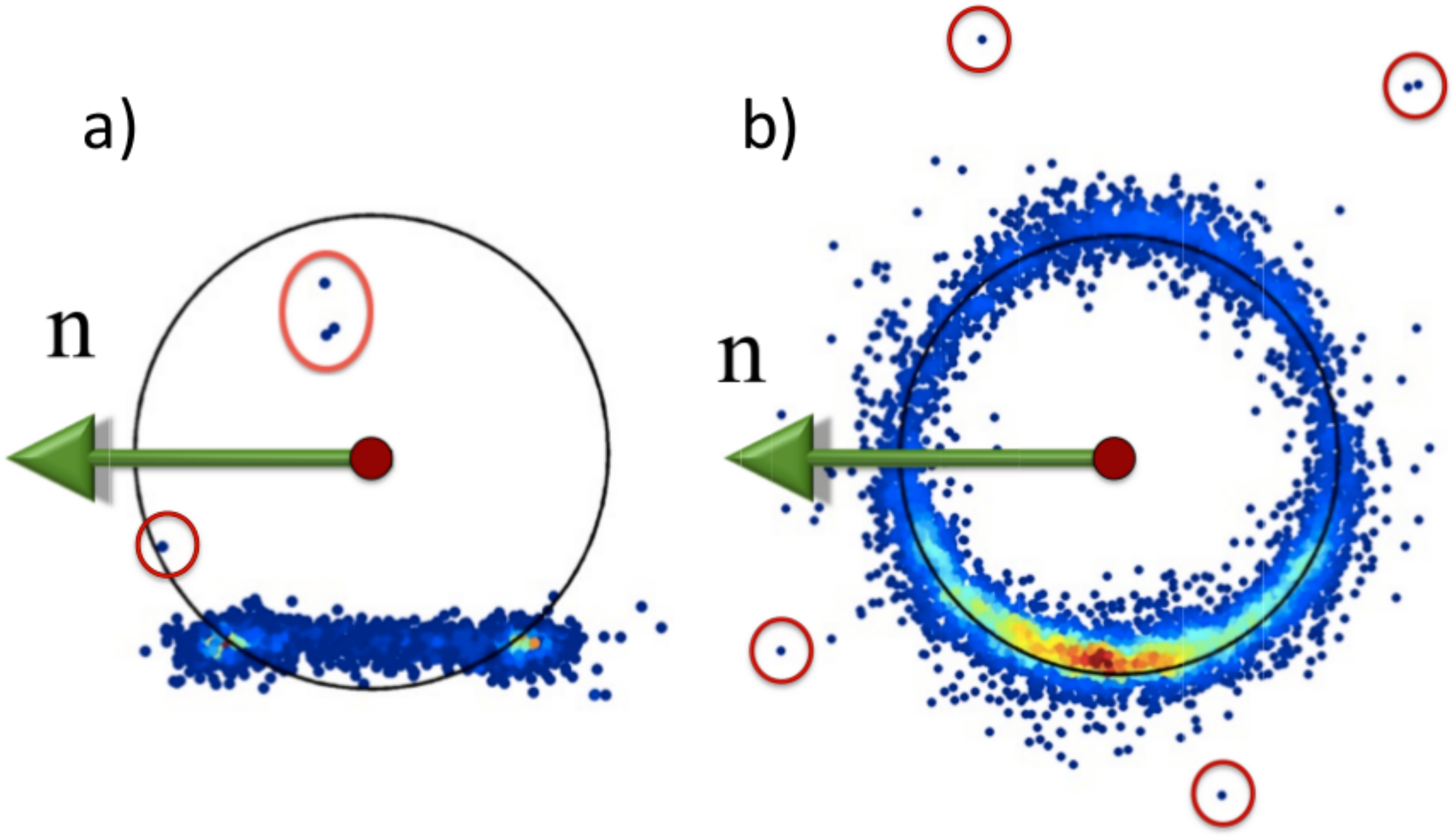}
        \caption{{ 
     (Color online) 
a) The proline contribution to the non-aromatic  level-$\delta$ atoms in Figure \ref{fig20} b). Three
apparent outliers have been encircled.  
%The (black) circle
%denotes the average distance of the distribution from the center.
b) The level-$\delta$ distribution of O atoms. The (black) circle
denotes the average distance of the distribution from the center.
Some outliers have been highlighters with red circles.
}}
       \label{fig21}
       \end{figure}
we have the proline contribution to figure \ref{fig20} b) and in 
figure \ref{fig21} b) we show the distribution of the O atoms at level-$\delta$. 
The latitude angles in O are highly restrained while the longitudinal angles 
are quite flexible. Some 
apparent outliers have been encircled in both figures \ref{fig21}, as generic examples. 

Finally, as in figure \ref{fig13} in figures \ref{fig20} and \ref{fig21} there is no visible sign of secondary structure:
The standard ${\mathcal X}_2$ dihedral is backbone independent. 

However, as in figures \ref{fig14},
in the backbone Frenet frames where the C$_\alpha$ is located at the 
center of the sphere, the secondary structure dependence becomes \emph{visible} in the level-$\delta$ rotamers. 
As an  example, we show in figure \ref{fig22} 
\begin{figure}[h]
        \centering
                \includegraphics[width=0.45\textwidth]{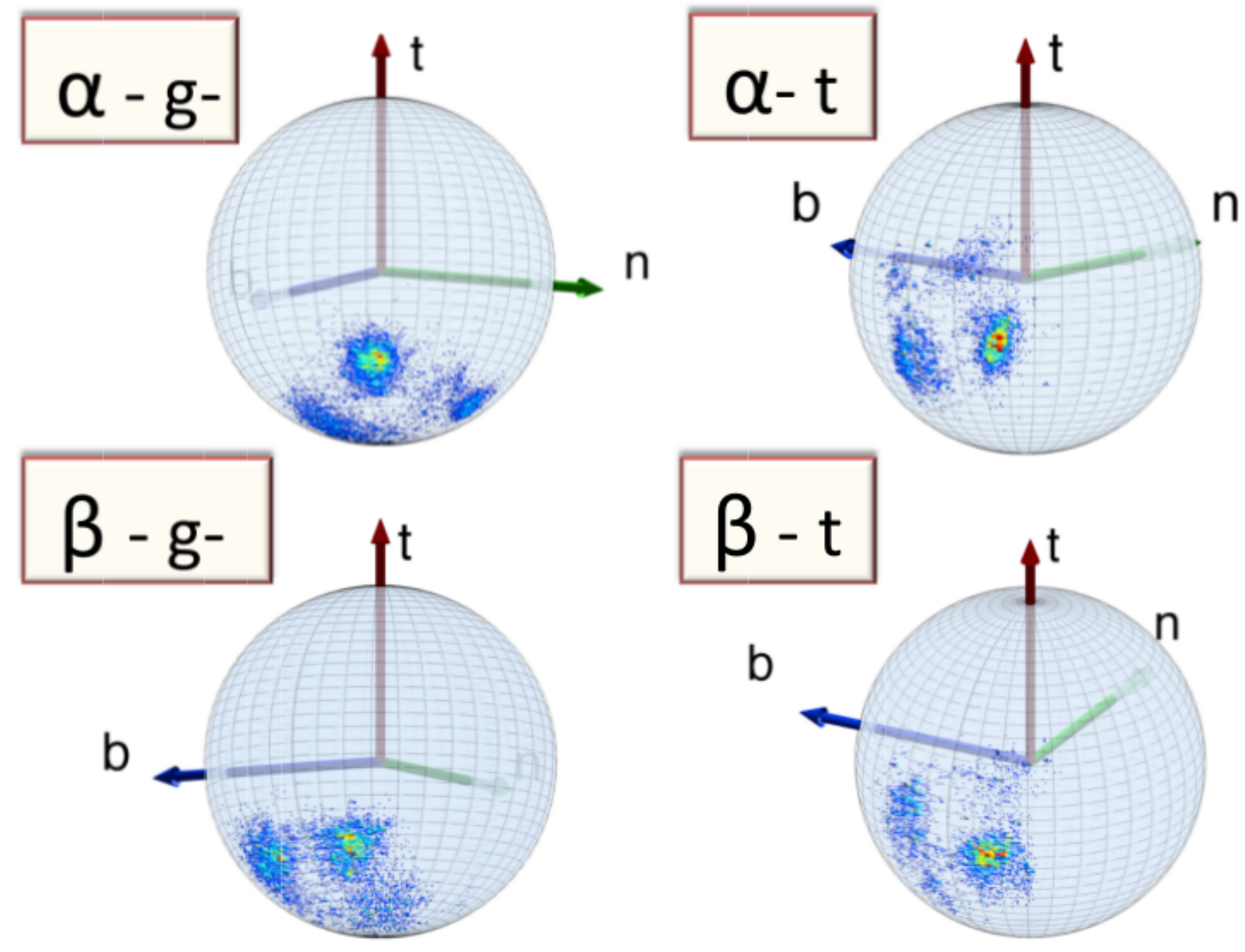}
        \caption{{ 
     (Color online) 
 Four figures that show the level-$\delta$ Frenet frame distributions corresponding to the level-$\gamma$
distributions in figures \ref{fig14}.  The labeling is as follows: 
$\alpha$ - $g$- stand for $\alpha$-helical backbone secondary structure in $g$- rotamer in figures \ref{fig14},
$\alpha$ - $t$ stand for $\alpha$-helical backbone secondary structure in $t$ rotamer in figures \ref{fig14},
$\beta$ - $g$- stand for $\beta$-stranded backbone secondary structure in $g$- rotamer in figures \ref{fig14}
and $\beta$ - $t$ stand for $\beta$-stranded  backbone secondary structure in $t$ rotamer in figures \ref{fig14}.
}}
       \label{fig22}
       \end{figure}
how some of the regions in figure \ref{fig14} are seen on the surface of the 
ensuing C$_\alpha$ centered sphere, by the roller coasting observer. 
The examples we have displayed are the overlap of the $\alpha$-helical
structures with the $g-$  rotamer (marked $\alpha$-$g-$ in the figure ) and $t$ rotamer ($\alpha$-$t$), and the 
overlap of the $\beta$-stranded
structures with the $g-$ rotamer ($\beta$-$g-$) and $t$ rotamer ($\beta$-$t$). A  secondary structure dependent
trimodal  rotamer structure is clearly present, in each of the distributions.

%
%
%%%%%%%%%%%%%%%%%%%%%%%%%%%%%%%%%%%%%%%%%%%%%%%%%%

\subsubsection{Secondary structure dependent level-$\delta$ rotamer angles:}

%
%
%%%%%%%%%%%%%%%%%%%%%%%%%%%%%%%%%%%%%%%%%%%%%%%%%%

Following (\ref{tbeta})-(\ref{bbeta}) and figures \ref{fig15}-\ref{fig17} we
proceed to visually inspect secondary structure dependence in
the level-$\delta$ rotamers. For this, we define an orthonormal frame as follows:
\[
{\mathbf t}_\gamma \ = \ \frac{ \mathbf r_{\gamma} - \mathbf r_{\beta} }
{| \, \mathbf r_{\gamma} - \mathbf r_{\beta}  \, |}
\]
\[
\mathbf n_\gamma \ = \ \frac{ \mathbf t_\gamma \times \mathbf t_{\alpha} }{ | \, \mathbf t_\gamma \times 
\mathbf t_{\alpha} \, |}
\]
Finally, 
\[
\mathbf b_\gamma = \mathbf t_\gamma \times \mathbf n_\gamma
\]

We start with the non-aromatic amino acids. In figure \ref{fig23} 
\begin{figure}[h]
        \centering
                \includegraphics[width=0.45\textwidth]{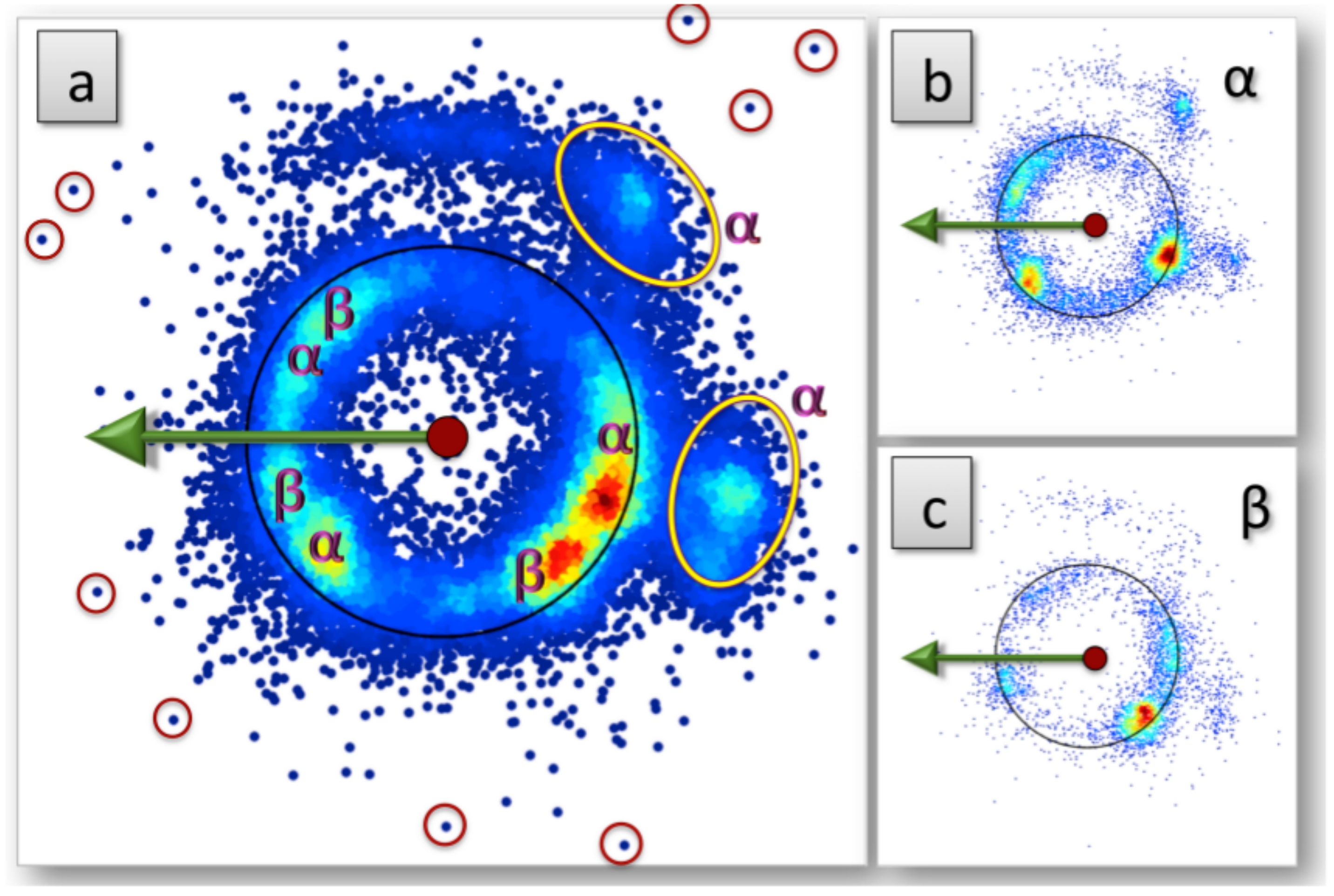}
        \caption{{ 
     (Color online) 
 The level-$\delta$ distribution of non-aromatic 
C atoms, in the stereographic projection. In figure a) we show the
entire background, and in b) and c) those that have been classified as
$\alpha$-helical and $\beta$-stranded, respectively. Some outliers have also been marked.
}}
       \label{fig23}
       \end{figure}
we show the distribution of all 
the C$_\delta$ non-aromatic atoms in our data set. In this figure we have also identified 
those  apparent rotamers that are classified either as $\alpha$-helical or $\beta$-stranded in 
PDB. The figure shows that there is a clear secondary structure dependence in these rotamers. 
In figure \ref{fig24} 
\begin{figure}[h]
        \centering
                \includegraphics[width=0.45\textwidth]{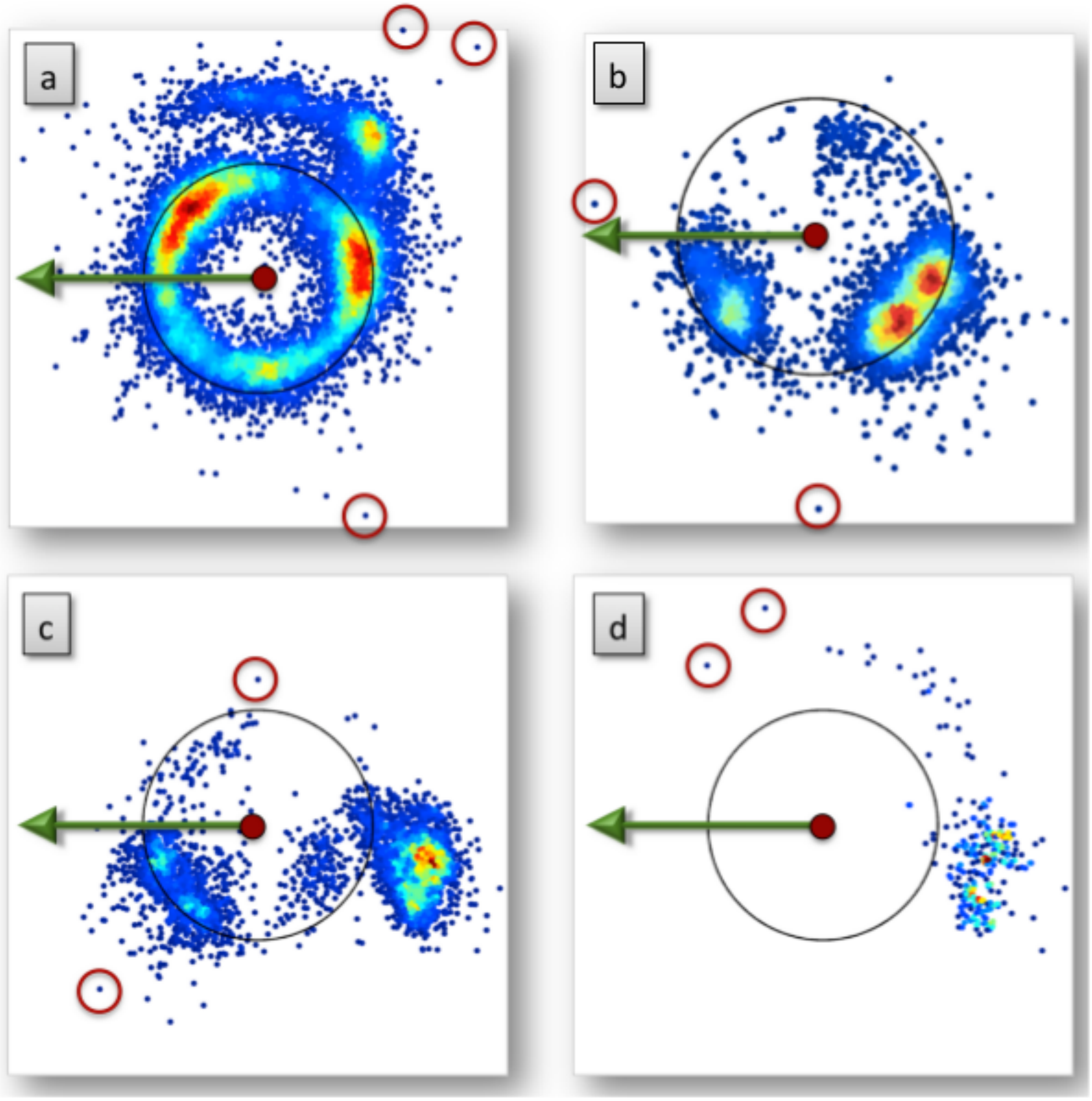}
        \caption{{ 
     (Color online) 
 The level-$\delta$ distribution of non-aromatic 
C atoms, in the stereographic projection and divided according to the level-$\gamma$ rotamers.
In figure a) we show the $g$- rotamer, in figure b) we show the $t$ rotamer, in figure c) we have the $g$+
rotamer and in figure d) we show the $cis$-proline. The radii of the (black) circles coincide with the average
latitude in figure \ref{fig23} a). Some outliers have also been encircled.
}}
       \label{fig24}
       \end{figure}
we display the three level-$\gamma$ subsets of \ref{fig23} a). Again, there is a
clear secondary structure dependence in the rotamers. We have also encircled some apparent outliers
in both figures \ref{fig23} and \ref{fig24}. Far-away outliers also exist  (not shown),
these can be located and visualized by rotating the original
sphere as in figure \ref{fig18}.
 
We proceed to the aromatic amino acids. In figure \ref{fig25} a) 
\begin{figure}[h]
        \centering
                \includegraphics[width=0.45\textwidth]{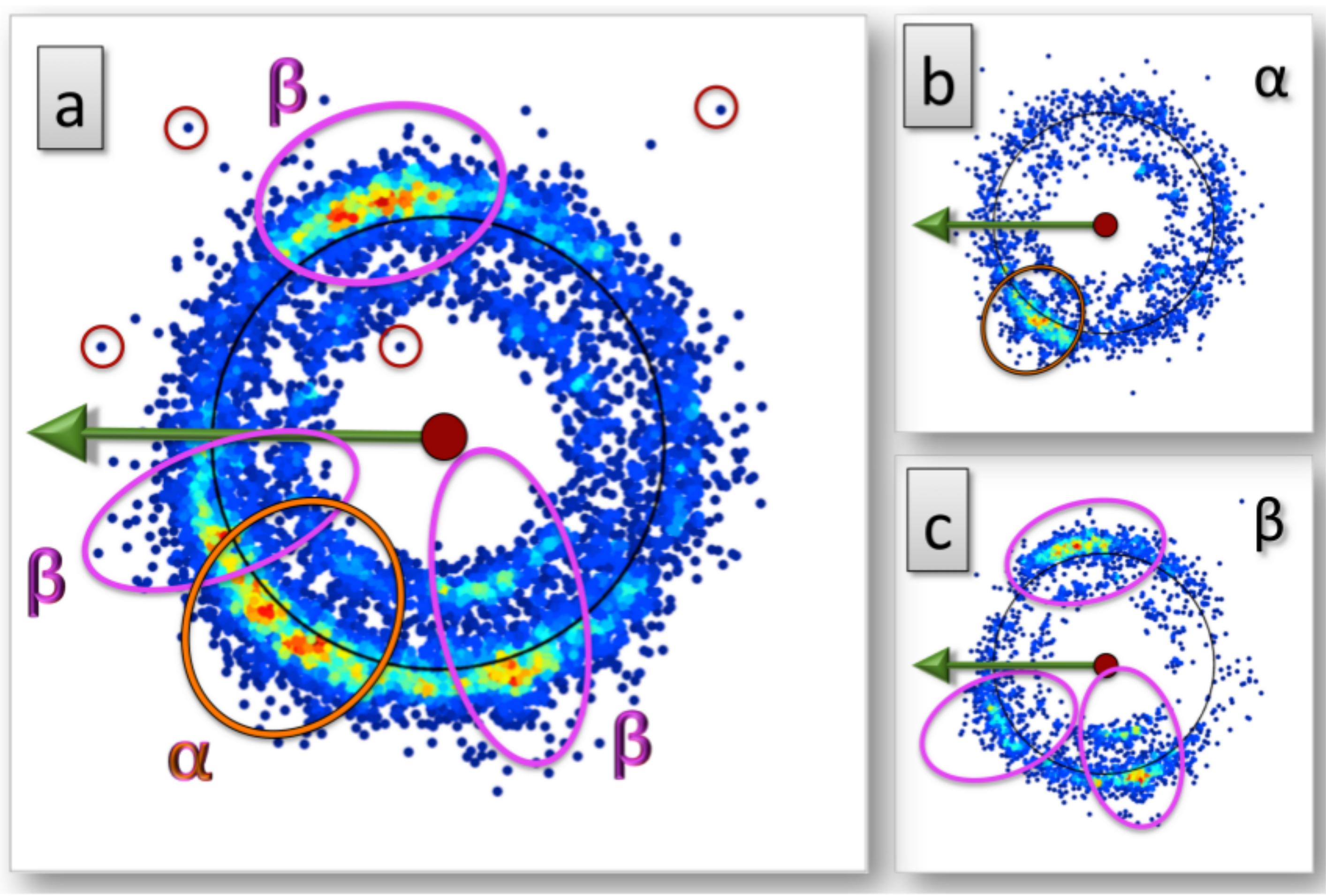}
        \caption{{ 
     (Color online) 
 Level-$\delta$ distribution of aromatic 
CD1 atoms PHE, TYR and TRP  in the stereographic projection. In a) all CD1 atoms in our data set,
and in b) and c) the $\alpha$-helical and $\beta$-stranded subsets, with rotamer states encircled . Some outliers have also been encircled in
a). 
}}
       \label{fig25}
       \end{figure}
we show all level-$\delta$ carbons (CD1 in PDB),
these are PHE, TYR, TRP. In figures \ref{fig25} b) and c) we show the subsets of \ref{fig25} a) that have been classified as 
$\alpha$-helical {\it resp.} $\beta$-stranded in PDB. 
In \ref{fig26} a) 
\begin{figure}[h]
        \centering
                \includegraphics[width=0.45\textwidth]{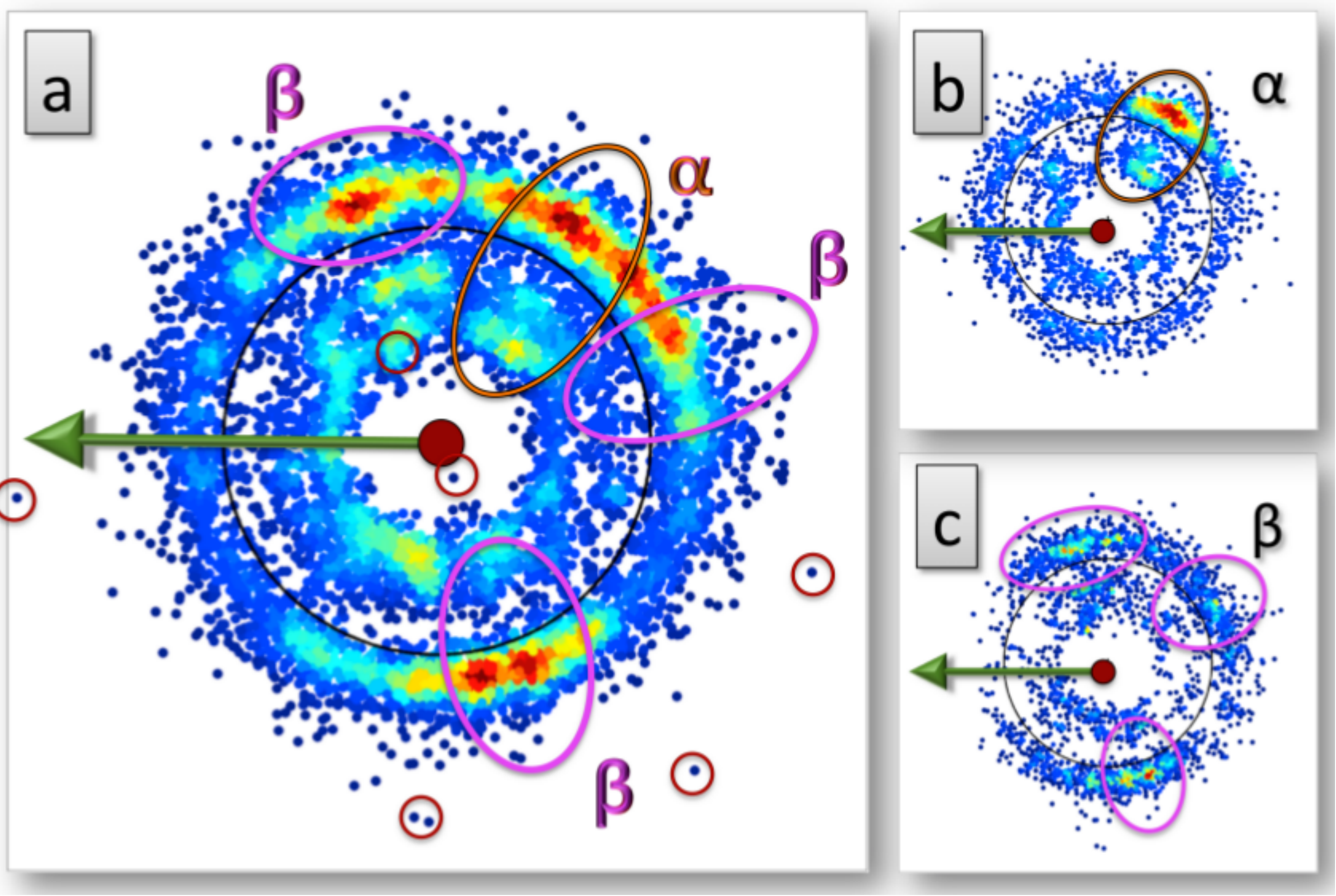}
        \caption{{ 
     (Color online) 
 Same as figure \ref{fig25}, but for the CD2 carbons according to PDB classification, 
including PHE, TYR, TRP  and HIS.
Some outliers have also been encircled in figure \ref{fig26} a).}}
       \label{fig26}
       \end{figure}
we show all level-$\delta$ carbons (CD2 in PDB) {\it i.e.} PHE, TYR, TRP and HIS.
In figures \ref{fig26} b) and c) we show the subsets of \ref{fig26} a) that have been classified as 
$\alpha$-helical {\it resp.} $\beta$-stranded in PDB. In both figures \ref{fig25} and \ref{fig26}
the secondary structure dependence is again manifest. In particular, both $\alpha$-helices and $\beta$-strands form
clear rotamers. 
We have also highlighted some outliers, by encircling them.

\subsection{Level$-\epsilon$ atoms}

%%%
%
%
%
%
%
%%%%%%%%%%%%%%%%%%%%%%%%%%%%

We proceed to the level-$\epsilon$ atoms. We follow the previous construction: We 
introduce a coordinate frame which is based 
at the C$_\delta$ carbon {\it i.e.} describes the point-of-view of an imaginary minuscule observer who has
climbed up to C$_\delta$ along the side chain. 
We map the level-$\epsilon$ atoms on the surface of the two-sphere 
which is centered at the C$_\delta$, followed by the stereographic projection. 

Note that in  the case of PHE and TYR two essentially identical choices can be made.
In the case of TRP there are also two choices, and we choose the one denoted CD2 in PDB, it is
covalently bonded to the higher level C atoms.
In the case of HIS a framing could also be based on the level-$\delta$ N atom, but 
here we select the level-$\delta$ C atoms that are denoted CD2 in PDB.

The orthonormal triplet is now defined as follows,
\[
{\mathbf t}_\delta \ = \ \frac{ \mathbf r_{\delta} - \mathbf r_{\gamma} }
{| \, \mathbf r_{\delta} - \mathbf r_{\gamma}  \, |}
\]
\[
\mathbf n_\delta \ = \ \frac{ \mathbf t_\delta \times \mathbf t_{\alpha} }{ | \, \mathbf t_\delta \times 
\mathbf t_{\alpha} \, |}
\]
and 
\[
\mathbf b_\delta = \mathbf t_\delta \times \mathbf n_\delta
\]

%
%
%
%
%
%
%       %
%
%
%

In figures \ref{fig27} a)-f)
\begin{figure}[h]
        \centering
                \includegraphics[width=0.45\textwidth]{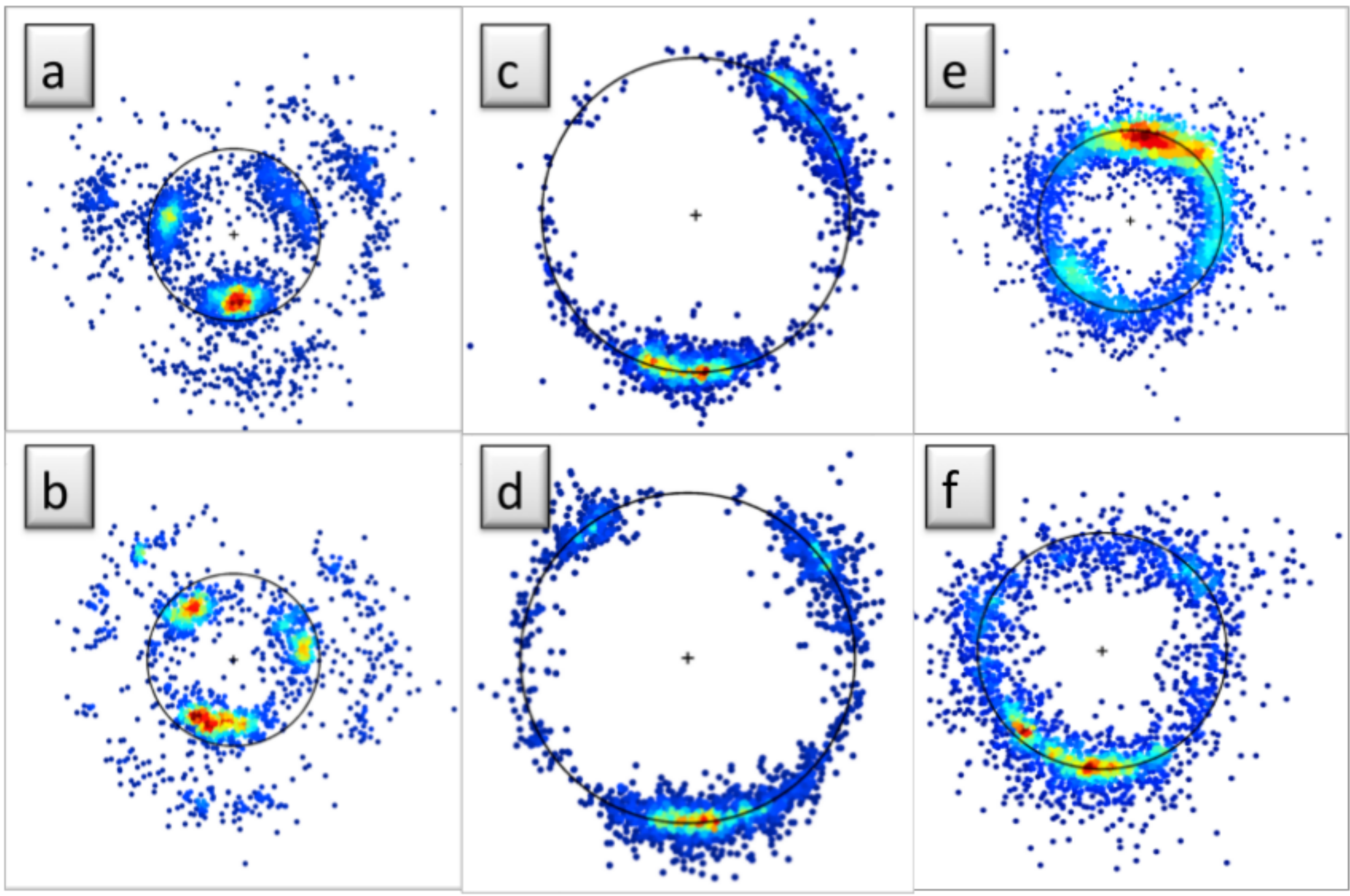}
        \caption{{ 
     (Color online) 
Examples of rotamers in level-$\epsilon$ atoms; the black circles have
the same radius in a) and b), in c) and d), and in e) and f).  
In figure a) the $\alpha$-helix and in b) the $\beta$-strand rotamers for
for CE in MET and LYS; the structures outside the
circle are LYS, those inside are MET. In figure c) the $\alpha$-helix and in d) the $\beta$-strand rotamers for
for CE1 in PHE and TYR. In figure e) the $\alpha$-helix rotamers for OE1 in GLU and GLN, and in f) 
the $\alpha$-helix rotamers for OE2 in GLU (there is no GLN). }}
       \label{fig27}
       \end{figure}
we show various examples of level-$\epsilon$ atoms. We observe that in addition of rotamers in the 
longitude, there
are also rotamer-like variations in the latitude angle, as shown in black circles in each figure.
%
%
%
%
%
%
%%%%%%%%%%%%%%%%%%%%%%%%%%%%%%%%%%%%%%%%%%%%%%%%%%
%
%%
%
%%%%%%%%%%%%%%%%%%%%%%%%%%%%%%%%%%%%%%%%%%%%%%%%%%%
%
%

\subsection{Level-$\zeta$ atoms}

We continue  to level-$\zeta$. We introduce the C$_\epsilon$ centered two-sphere with orthonormal
triplet given by 
\[
{\mathbf t}_\epsilon \ = \ \frac{ \mathbf r_{\epsilon} - \mathbf r_{\delta} }
{| \, \mathbf r_{\epsilon} - \mathbf r_{\delta}  \, |}
\]
\[
\mathbf n_\epsilon \ = \ \frac{ \mathbf t_\epsilon \times \mathbf t_{\alpha} }{ | \, \mathbf t_\epsilon \times 
\mathbf t_{\alpha} \, |}
\]
\[
\mathbf b_\epsilon = \mathbf t_\epsilon \times \mathbf n_\epsilon
\]

As an example, in figures \ref{fig28} 
\begin{figure}[h]
        \centering
                \includegraphics[width=0.45\textwidth]{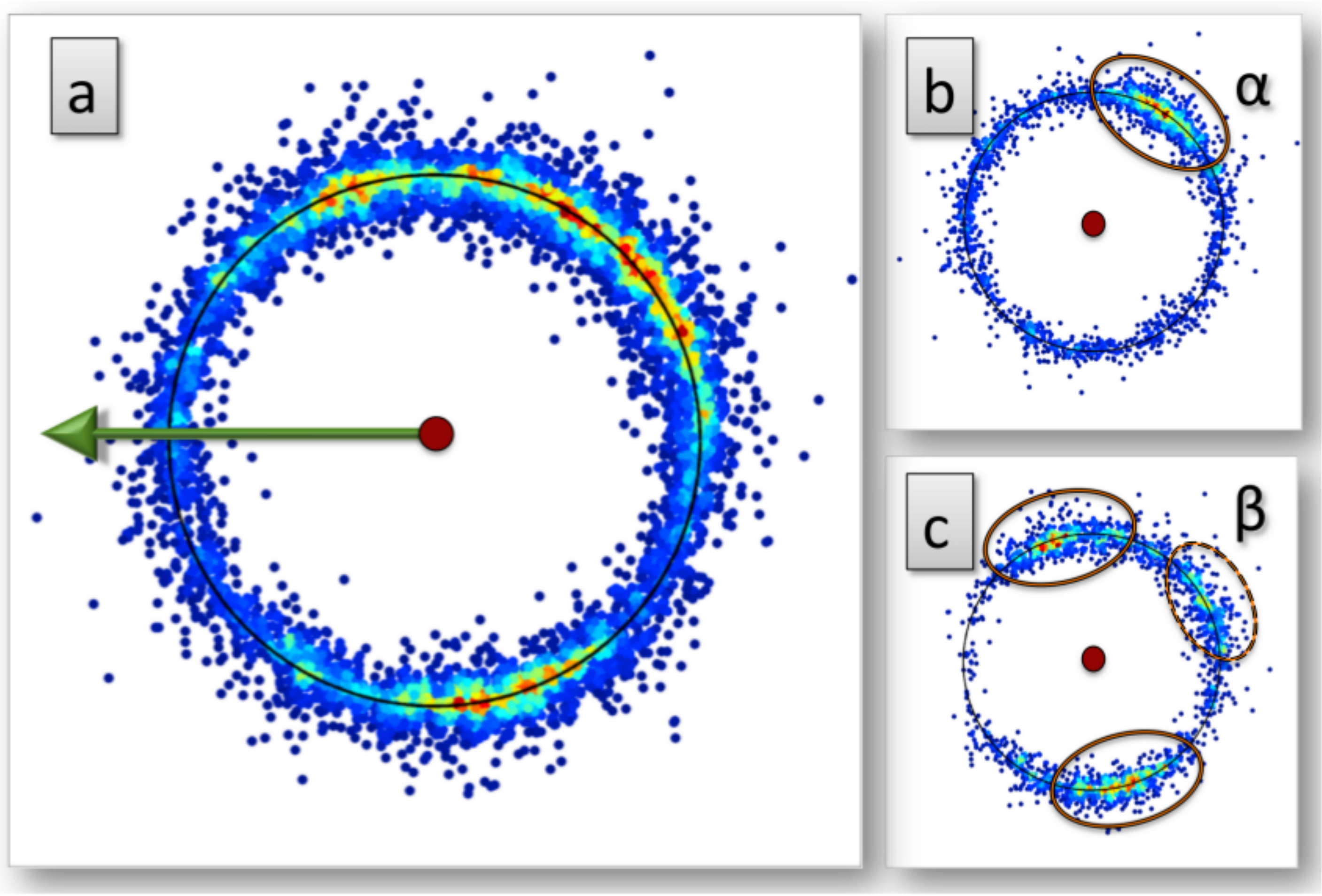}
        \caption{{ 
     (Color online) 
 Example of level-$\zeta$ rotamers. In figure a) we have all the C$_\zeta$ carbons in PHE and TYR. In
figures b) and c) we show the subsets that correspond to $\alpha$-helical and $\beta$-stranded secondary structures,
respectively. }}
       \label{fig28}
       \end{figure}
we show the  C$_\zeta$ carbons for PHE and TYR using stereographic
projection. The figure \ref{fig28} a) shows all C$_\zeta$ atoms, and figures \ref{fig28} b) and c) show the 
$\alpha$-helical and $\beta$-stranded subsets. In the case of $\alpha$-helical secondary structures
we identify one rotamer. In the case of $\beta$-stranded structures we observe three rotamers.
We observe that the $\beta$-stranded rotamers are not distributed evenly. The rotamers are
not related to each other by (regular) $120^{\rm o}$ rotations.

%
%
%
%
%
%
%
%%%%%%%%%%%%%%%%%%%%%%%%%%%%%%%%%%%%%%%%%%%%%%%%%%%
%
%
%%
%%
%%%%%%%%%%%%%%%%%%%%%%%%%%%%%%%%%%%%%%%%%%%%%%%%%%
%
%
%%%
%
%
%
%

\subsection{Level-$\eta$ atoms}

We continue the process to the  level-$\eta$ which is the final level in proteins.  We follow our construction
to define the C$_\zeta$ centered coordinate system, with 
\[
{\mathbf t}_\zeta \ = \ \frac{ \mathbf r_{\zeta} - \mathbf r_{\epsilon} }
{| \, \mathbf r_{\zeta} - \mathbf r_{\epsilon}  \, |}
\]
\[
\mathbf n_\zeta \ = \ \frac{ \mathbf t_\zeta \times \mathbf t_{\alpha} }{ | \, \mathbf t_\zeta \times 
\mathbf t_{\alpha} \, |}
\]
\[
\mathbf b_\zeta = \mathbf t_\zeta \times \mathbf n_\zeta
\]
As before, we also introduce the ensuing stereographic projection. 

As an example, in figures \ref{fig29} 
\begin{figure}[h]
        \centering
                \includegraphics[width=0.45\textwidth]{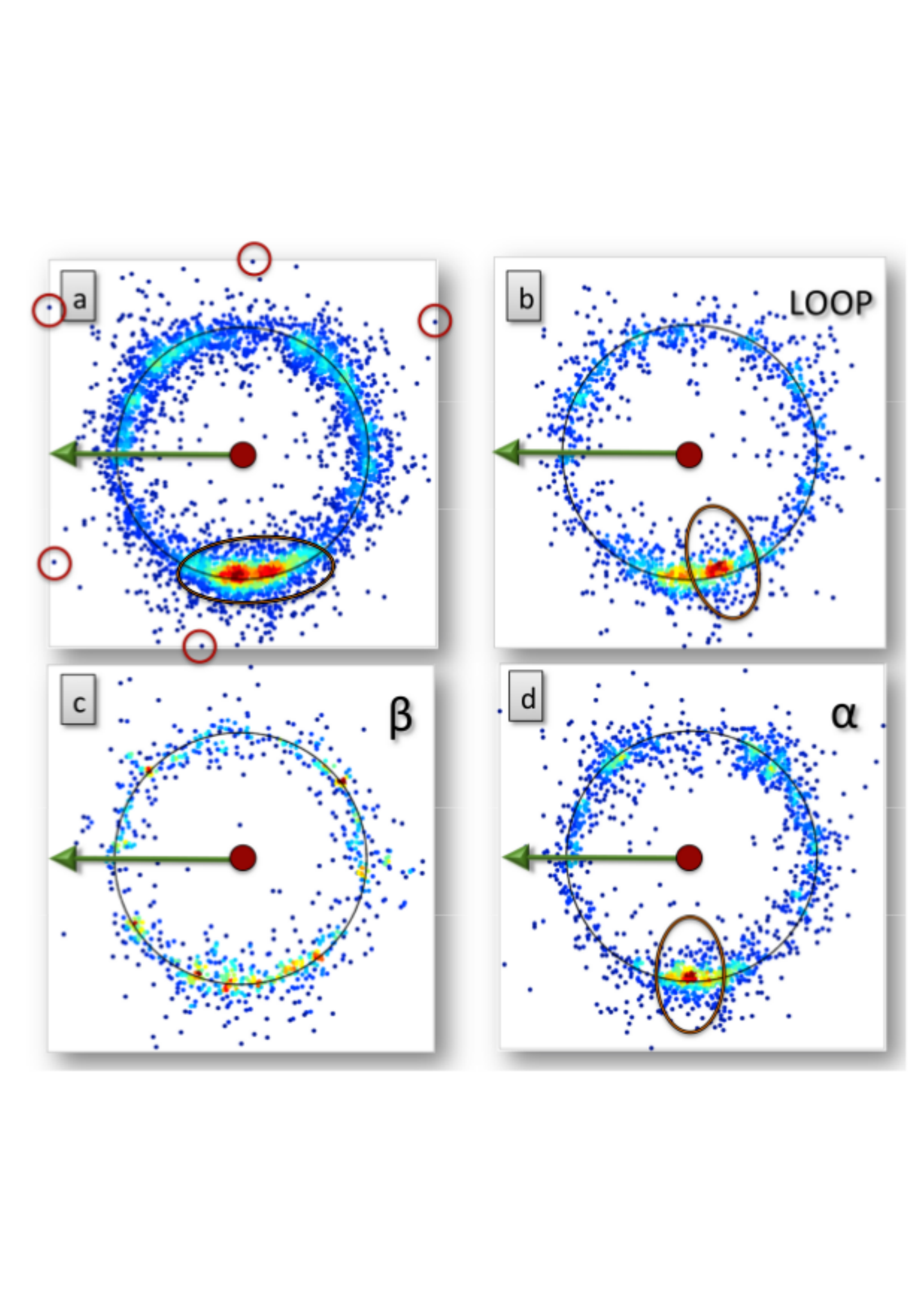}
        \caption{{ 
     (Color online) 
 Example of level-$\eta$ rotamers. In figure a) we have all the N$\eta 2$ atoms in ARG.
There are two very close rotamer states, that have been encircled. Some outliers have also been
encircled.
In figures b)-d) we show the subsets that correspond to loops, $\beta$-stranded and $\alpha$-helical
secondary structures,
respectively. Comparison of the figures reveals that the two very close-by
rotamers in a) correspond to loops and $\alpha$-helices.}}
       \label{fig29}
       \end{figure}
we display the N${\eta}2$ distribution in ARG.  
Now there is a very strong two-fold localization of the distribution, shown in figure \ref{fig29} a).
In figures b)-d) we consider the subsets, consisting of PDB secondary structures that are classified
as loops b), $\beta$-strands c) and $\alpha$-helices d). These identify the two rotamers
in figure \ref{fig29} a). Some of the outliers are encircled, as examples, in a).

%
%
%
%%
%%
%
%
%%%
%%%%%%%%%%%%%%%%%%%%%%%%%%%%%%%%%%%%%%%%%%%%%%%%%%%
%
%
%
%

%
%
%
%
%%
%
%
%
%
%
%
%
%
%
%
%
%
%
%
%
%
%
%
%
%
%
%
%
%
%
%
%
%
%

\section{Discussion}
We have utilized recent developments in modern 3D visualization techniques and advances in
virtual reality
to describe how to construct an entirely C$_\alpha$ geometry based visual library of the backbone 
and side chain atoms. Our construction is based on progress in 
visualization  that  has taken place since the inception 
of the Ramachandran map. 
In lieu of a torus, our approach engages the geometry of a sphere and as such it has a 
direct "what-you-see-is-what-you-have" visual correspondence
to the protein structure. In particular,
we utilize the geometrically determined discrete 
Frenet frames of  \cite{Hu-2011}. We propose the concept of 
an imaginary observer, chosen so that the discrete 
Frenet frames determine the orientation of the observer when she
roller-coasts  along the backbone and climbs up the side chains. She maps the directions of all the heavy 
atoms on the surface of a two-sphere that  surrounds her, exactly as these atoms are seen in her local frame like
stars in the sky.   

Since the discrete Frenet frames can be unambiguously determined in terms of the C$_\alpha$ 
trace only, we can analyze both the backbone atoms and the side chain atoms on 
equal footing, in a single geometric framework. This is not possible
in the conventional Ramachandran approach, that 
assumes {\it a priori} knowledge of the peptide planes, to define the dihedral angles.

As examples of the approach, we 
have analyzed the orientation of various heavy atoms that are 
located both along the backbone and in
the side chains. Our approach also enables a direct, {\it visual } identification
of outliers. 

In particular, we have found that in terms of the discrete Frenet frames, the secondary
structure dependence becomes clearly visible in the rotamer structure, both in the case of
the backbone atoms and in the case of the side chain atoms. Apparently this is not always the 
case, in conventional approaches such as \cite{Holm-1991,Rotkiewicz-2008,Li-2009};
according to \cite{Dunbrack-2002} conventional secondary structure dependent rotamer libraries
do not provide much more  information than backbone-independent rotamer libraries.  
But by using the Frenet frame coordinate system chosen here, there is a clear correlation between
secondary structures and rotamer positions.  Thus the approach we have presented,   
can form a basis for the future development of a novel approach to the C$_\alpha$ trace problem. 
Unlike the existing approaches \cite{Holm-1991,Rotkiewicz-2008,Li-2009} the one we envision
accounts for the secondary structure dependence in the heavy atom positions that we have revealed,
which should lead to an improved accuracy  in determining the heavy atom positions.

\vskip 0.4cm
\begin{flushleft}
{\bf Acknowledgements} 
\end{flushleft}

A.J. Niemi thanks A. Elofsson, J. Lee and A. Liwo for a discussion. This research hs been
supported by a CNRS PEPS Grant, Region Centre Rech\-erche d$^{\prime}$Initiative Academique grant,
Cai Yuanpei Exchange Program, Qian Ren Grant at BIT, Carl Trygger's Stiftelse f\"or vetenskaplig forskning, and Vetenskapsr\aa det.

\end{document}